\title{
Parameterized Pattern Matching -- Succinctly
}

\author{
Arnab Ganguly\thanks{
School of Electrical Engineering and Computer Science. 
Louisiana State University, Baton Rouge, USA.
\textbf{email}: agangu4@lsu.edu}
\and
Rahul Shah\thanks{
School of Electrical Engineering and Computer Science. 
Louisiana State University, Baton Rouge, USA.
Also affiliated to: National Science Foundation, USA.
\textbf{email}: rahul@csc.lsu.edu, rahul@nsf.gov}
\and 
Sharma V. Thankachan\thanks{
School of Computational Science and Engineering. 
Georgia Institute of Technology, Atlanta, USA.
\textbf{email}: sharma.thankachan@gatech.edu}
}

\date{\today}

\documentclass[11pt]{article}

\usepackage{graphicx,amsmath,amssymb}
\usepackage{algorithm}
\usepackage{algpseudocode}
\usepackage{geometry}
\usepackage{pdflscape}
\usepackage{subcaption}
\usepackage{tabularx}
\usepackage{enumitem}
\usepackage{hyperref}

\newcommand{\calP}{\mathcal{P}}

\newcommand{\Z}{\mathsf{Z}}

\newcommand{\s}{\mathcal{S}}

\newcommand{\T}{\mathsf{T}}
\newcommand{\calT}{\mathcal{T}}
\newcommand{\calTRev}{\overleftarrow{\calT}}
\newcommand{\calD}{\mathcal{D}}
\renewcommand{\P}{\mathsf{P}}

\newtheorem{lem}{Lemma} 
\newcommand {\BL} {\begin{lem}} 
\newcommand {\EL} {\end{lem}} 

\newtheorem{cor}  {Corollary} 
\newcommand {\BCR} {\begin{cor}}
\newcommand {\ECR} {\end{cor}}

\newtheorem{sub} {Subquery} 
\newcommand {\BSQ} {\begin{sub}}
\newcommand {\ESQ} {\end{sub}}

\newtheorem{thm} {Theorem} 
\newcommand {\BT} {\begin{thm}}
\newcommand {\ET} {\end{thm}}

\newtheorem{prob} {Problem} 
\newcommand {\BDE} {\begin{prob}}
\newcommand {\EDE} {\end{prob}}

\newtheorem{defi} {Definition} 
\newcommand {\BD} {\begin{defi}}
\newcommand {\ED} {\end{defi}}

\newtheorem{obs} {Observation} 
\newcommand {\BO} {\begin{obs}}
\newcommand {\EO} {\end{obs}}

\newtheorem{fact}  {Fact} 
\newcommand {\BF} {\begin{fact}} 
\newcommand {\EF} {\end{fact}} 

\newtheorem{convention}  {Convention} 
\newcommand {\BC} {\begin{convention}} 
\newcommand {\EC} {\end{convention}} 

\newtheorem{assum}  {Assumption} 
\newcommand {\BA} {\begin{assum}} 
\newcommand {\EA} {\end{assum}} 

\newenvironment{proof}{\textbf{Proof:}}

\newcommand{\ST}{\mathsf{ST}}
\newcommand{\pST}{\mathsf{pST}}
\newcommand{\pSA}{\mathsf{pSA}}
\newcommand{\pSAI}{\mathsf{pSA^{-1}}}

\newcommand{\SA}{\mathsf{SA}}
\newcommand{\SAI}{\mathsf{SA^{-1}}}
\newcommand{\LF}{\mathsf{LF}}

\newcommand{\tST}{\mathsf{sST}}
\newcommand{\tSA}{\mathsf{sSA}}
\newcommand{\tSAI}{\mathsf{sSA^{-1}}}

\renewcommand{\O}{\mathcal{O}}
\newcommand{\BWT}{\mathsf{BWT}}
\newcommand{\pLF}{\mathsf{pLF}}
\newcommand{\tLF}{\mathsf{sLF}}
\newcommand{\pBWT}{\mathsf{pBWT}}
\newcommand{\tBWT}{\mathsf{sBWT}}

\newcommand{\rangeCount}{\mathsf{rangeCount}}
\newcommand{\rank}{\mathsf{rank}}
\newcommand{\select}{\mathsf{select}}
\newcommand{\pred}{\mathsf{predecessor}}
\newcommand{\rangeNV}{\mathsf{rangeNextVal}}
\newcommand{\rangePV}{\mathsf{rangePrevVal}}

\newcommand{\leafc}{\mathsf{leafLeadChar}}
\newcommand{\pc}{\mathsf{pCount}}
\newcommand{\prev}{\mathsf{prev}}
\newcommand{\prevRev}{\mathsf{\overleftarrow{\prev}}}
\newcommand{\compl}{\mathsf{compl}}
\newcommand{\CountL}{\mathsf{Count_<}}
\newcommand{\CountE}{\mathsf{Count_=}}

\newcommand{\lca}{\mathsf{lca}}
\newcommand{\parent}{\mathsf{parent}}
\newcommand{\child}{\mathsf{child}}

\newcommand{\lmostLeaf}{\mathsf{lmostLeaf}}
\newcommand{\rmostLeaf}{\mathsf{rmostLeaf}}
\newcommand{\la}{\mathsf{levelAncestor}}

\newcommand{\zeroNode}{\mathsf{zeroNode}}
\newcommand{\nd}{\mathsf{nodeDepth}}
\newcommand{\alphaDepth}{\mathsf{zeroDepth}}
\newcommand{\fc}{\mathsf{fCount}}
\newcommand{\fs}{\mathsf{fSum}}

\newcommand{\zeroNodePos}{\zeroNode^+}
\newcommand{\fcPos}{\fc^+}
\newcommand{\fsPos}{\fs^+}
\newcommand{\zeroNodeNeg}{\zeroNode^-}
\newcommand{\fcNeg}{\fc^-}

\newcommand{\fsNegRev}{\overleftarrow{\fs}^-}

\newcommand{\pre}{\mathsf{pre\mbox{-}order}}

\newcommand{\nodePath}{\mathsf{path}}
\newcommand{\nodePathRev}{\mathsf{\overleftarrow{\nodePath}}}

\newcommand{\Leaf}{\mathsf{leaf}}
\renewcommand{\sp}{\mathsf{sp}}
\newcommand{\ep}{\mathsf{ep}}

\newcommand{\nxt}{\mathsf{next}}
\newcommand{\failure}{\mathsf{failure}}
\newcommand{\report}{\mathsf{report}}

\newcommand{\qed}{\hfill$\blacksquare$}

\begin{document}
\maketitle

\begin{abstract}
The fields of succinct data structures and compressed text indexing have seen quite a bit of progress over the last 15 years. 
An important achievement, primarily using techniques based on the Burrows-Wheeler Transform (BWT), was obtaining the full functionality of the suffix tree in the optimal number of bits.
A crucial property that allows the use of BWT for designing compressed indexes is \emph{order-preserving suffix links}. 
Specifically, the relative order between two suffixes in the subtree of an internal node is same as that of the suffixes obtained by truncating the first character of the two suffixes. 
Unfortunately, in many variants of the text-indexing problem, for e.g., parameterized pattern matching and 2D pattern matching, this property does not hold.
Consequently, the compressed indexes based on BWT do not directly apply. 
Furthermore, a compressed index for any of these variants has been elusive throughout the advancement of the field of succinct data structures. 
We achieve a positive breakthrough on one such problem, namely the \emph{Parameterized Pattern Matching} problem.

Let $\T$ be a text that contains $n$ characters from an alphabet $\Sigma$, which is the union of two disjoint sets:  $\Sigma_s$ containing static characters (s-characters) and $\Sigma_p$ containing parameterized characters (p-characters). 
A pattern $P$ (also over $\Sigma$) matches an equal-length substring $S$ of $\T$ iff the s-characters match exactly, and there exists a one-to-one function that renames the p-characters in $S$ to that in $P$.
The task is to find the starting positions (occurrences) of all such substrings $S$.
Previous indexing solution [Baker, STOC 1993], known as \emph{Parameterized  Suffix Tree}, requires $\Theta(n\log n)$ bits of space, and can find all $occ$ occurrences in time $\O(|P|\log \sigma+ occ)$, where $\sigma = |\Sigma|$.
In this paper, we present the first succinct index for this problem. 
Specifically, we introduce two indexes with the following space-and-time tradeoffs:
\begin{itemize}
\item $\O(n\log\sigma)$ bits and $\O(|P|\log \sigma+ occ\cdot  \log n)$ query time.
\item $n \log \sigma + \O(n)$ bits and $\O((|P|+ occ\cdot  \log n) \log\sigma\log \log \sigma)$ query time.
\end{itemize}
Furthermore, the techniques are extended to obtain the first succinct representation of the index of Shibuya for \emph{Structural Matching} [SWAT, 2000], and of Idury and Sch{\"{a}}ffer for \emph{Parameterized Dictionary Matching} [CPM, 1994].
\\ \\
{\bf Keywords:} Succinct data structures, Suffix trees, Burrows-Wheeler Transform
\end{abstract}

\newpage

\section{Introduction}
Pattern Matching is the algorithmic framework of finding all starting positions (or simply, occurrences) of a pattern in a text. 
In earlier works~\cite{BoyerM77,KarpR87,KnuthMP77}, both text and pattern were provided as input. 
In most cases, however, the text does not change and patterns come in as online query. 
This motivated the development of full text indexes i.e., pre-process the text and then build a data structure. 
Using the data structure, whenever a pattern comes as query, we can efficiently report all occurrences without having to read the entire text. 
Here and henceforth, $\T$ is a text containing $n$ characters chosen from an alphabet $\Sigma$ of size $\sigma \geq 2$, $P$ is a pattern containing $|P|$ characters (also from $\Sigma$), and $occ$ is the number of occurrences of $P$ in $\T$.
\emph{Suffix Trees}~\cite{Ukkonen:STC,Weiner:LPM} and \emph{Suffix Arrays} along with \emph{Longest Common Prefix} (LCP) array~\cite{Manber:SA} can answer pattern matching queries in $\O(|P| + occ)$ and in $\O(|P| + \log n + occ)$
time respectively. 
Both suffix trees and suffix arrays require $\Theta(n \log n)$ bits of space,  which is too large for most practical purposes (15-50 times the text). 
Grossi and Vitter~\cite{Grossi:2000:CSA}, and Ferragina and Manzini~\cite{FerraginaM00} were the first to address this problem.
Their respective space efficient indexes, namely \emph{Compressed Suffix Arrays} (CSA) and \emph{FM-Index}, have found a lot of applications, and led to the establishment of the field of compressed text indexing.
See~\cite{Navarro:2007:CFI} for a comprehensive survey.

Suffix trees have myriad applications (see Dan Gusfield's book~\cite{Gusfield1997}). 
For many applications, suffix trees have to be augmented with additional data.
After the initial advent of compressed text indexing structures, \emph{compressed suffix trees}~\cite{RussoNO11,Sadakane07} replaced the older suffix tree as a black-box. 
The focus was now on how to compress the augmenting data in informational theoretic minimal space.
Probably the first such work was that by Sadakane~\cite{SadakaneSODA02} who showed that given a (compressed) suffix array, the LCP array can be maintained in an additional $2n+o(n)$ bits.
Fischer~\cite{Fischer10} improved this to $o(n)$ bits.
The \emph{Range Minimum Query} (RMQ) data structure is an integral component to many applications related to suffix trees/arrays.
Fischer and Heun~\cite{FischerH07} showed how to maintain the RMQ data structure in $2n + o(n)$ bits.
In another direction, Sadakane~\cite{Sadakane02} considered the problem of document retrieval, and presented succinct indexes.
Later, other improvements (see~\cite{Navarro13} for a survey) culminated in the best known document listing index of Tsur~\cite{Tsur13}.
Other problems such as property matching~\cite{HonPST13} and matching statistics~\cite{OhlebuschGK10} also fall under the same theme.

To the best of our knowledge, no progress has been achieved on designing space efficient indexes (analogous to that of the Compressed Suffix Array and FM-index) for addressing the second kind of applications~\cite{Baker93,Giancarlo93,KimEFHIPPT14,Shibuya00} which require variants of the classical suffix tree.
We now formally define such a problem, that of parameterized pattern matching.

\BDE[\cite{Baker93}]
\label{problem}
Let $\Sigma$ be an alphabet of size $\sigma \geq 2$, which is the union of two disjoint sets: $\Sigma_s$ having $\sigma_s$ static characters (s-characters) and $\Sigma_p$ having $\sigma_p$ parameterized characters (p-characters). 
Two strings, having equal length, are a \emph{parameterized match} (p-match) if one can be transformed to the other by applying a one-to-one function that renames the characters in $\Sigma_p$.
Let $\T$ be a text having $n$ characters chosen from $\Sigma$. 
We assume that $\T$ terminates in an s-character $\$$ that appears only once.
The task is to index $\T$, such that for a pattern $P$ (also over $\Sigma$), we can report all  starting positions (occurrences) of the substrings of $\T$ that are a p-match with $P$.
\EDE 
Consider the following example. 
Let $\Sigma_s = \{A,B,C, \$\}$ and $\Sigma_p = \{w,x,y,z\}$. 
Then, $P = AxByCx$ p-matches within $\T[1,21] = AyBxCy AwBxCz xy AzBwCz\$$ at positions $1$ and $15$. 
At position $1$, the mapping is $x \rightarrow y$ and $y \rightarrow x$; whereas at position $15$, the mapping is $x \rightarrow z$ and $y \rightarrow w$. Note that $P$ does not match at position $7$ because $x$ would have to match with both $w$ and $z$.

\subsection{Applications}

Baker~\cite{Baker93} introduced this problem, with the main motivation being software plagiarism detection. 
They introduced a $\Theta(n \log n)$-bit index, known as the \emph{Parameterized Suffix Tree} (p-suffix tree), that can report all $occ$ p-matches in $\O(|P|\log\sigma+occ)$ time.
The role of the problem and the efficiency of p-suffix trees was presented using a program called \emph{Dup}~\cite{Baker95}.
Subsequently, the methodology became an integral part of various tools for software version management and clone detection, where identifiers and/or literals are renamed. 
The most state-of-the-art approaches~\cite{BaxterYMSB98,KoschkeFF06,RysselbergheD04} to clone detection use a hybrid approach, such as a combination of (i) a parse tree, which converts literals into parameterized symbols, and (ii) a p-suffix tree on top of these symbols. 
Unfortunately, as with traditional suffix trees, the space occupied by p-suffix trees is too large for most practical purposes, bringing in the demand for space-efficient variants.
In fact, one of the available tools (CLICS~\cite{CLICS}) very clearly acknowledges that the major space consumption is due to the use of p-suffix tree. 
Some other tools~\cite{Hummel5609665} use more IR type methodology for indexing repositories based on variants of the inverted index. 
Although less space consuming, there are no theoretical guarantees possible on query-times in such indexes. 
Since p-suffix trees accommodate substring match, tools based on them can also be powerful in recognizing smaller code fragment clones.
Following are a few other works that have used p-suffix trees: finding relevant information based on regular expressions in sequence databases~\cite{MouzaRS05,MouzaRS07}, detecting cloned web pages~\cite{di2001clone}, detecting similarities in JAVA sources from bytecodes~\cite{baker1998deducing}, etc.
Further generalizations of p-matching have also played an important role in computational biology for finding similar sequences~\cite{AmirACLP03,Shibuya00}.

\subsection{Contribution}
An important ingredient of suffix trees, crucial to compressed text indexing, is \emph{suffix links}. 
Specifically, let $u$ be a non-root node in the suffix tree, and let the string corresponding to the path from root to $u$ be $S$. 
Then, the node whose path label is obtained by truncating the first character of $S$ necessarily exists. 
Let $v$ be that node. 
The suffix link of $u$ points to $v$.
In suffix trees, the leaves are arranged in the lexicographic order of the suffix they represent.
Suffix links have the following so called \emph{order-preserving} property.
The leaves obtained by following suffix links from the leaves in $u$'s subtree appear in the same relative order in the subtree of $v$. 
Thus, the permutation of suffixes in $v$'s subtree can be encoded in terms of the permutation in $u$'s subtree. 
In applications like p-suffix tree~\cite{Baker93}, 2D suffix tree~\cite{Giancarlo93,KimKP03}, structural suffix tree~\cite{Shibuya00} etc., this property is not true. 
This brings in new challenges in how to encode such permutations, and whether succinct data structures are even possible. 
By presenting the first succinct index for Problem~\ref{problem}, we affirmatively answer this in the case of p-pattern matching.
The following theorem summarizes our main contribution.
\BT
\label{thm:param}
All $occ$ positions where $P$ is a p-match with $\T$ can be found as follows:
\begin{enumerate}[label={\emph{(\alph*)}},ref={\thethm\alph*}]
\item \label{thm:param:a} in $\O(|P|\log \sigma + occ\cdot \log n)$ time using an $\O(n\log \sigma)$-bit index.
\item \label{thm:param:b} in $\O((|P| + occ\cdot \log n) \log \sigma \log \log \sigma)$ time using an $n\log \sigma + \O(n)$-bit index.
\end{enumerate}
\ET
We introduce a BWT-like transform for a parameterized text, called the \emph{Parameterized} BWT. 
Using this, we show how to handle the \emph{order-inversion} in the case of p-suffixes when their first characters are truncated.
To achieve this, we implement analogous versions of the last-to-first column mapping and backward-search technique of Ferragina and Manzini~\cite{FerraginaM00}. 
This encompasses the major contribution of our work, which is achieved using newly introduced concepts, and existing succinct data structure toolkit.
The techniques are then extended to address two related problems.
Specifically, we show that an index with the same space-and-time tradeoff (as in Theorem~\ref{thm:param}) can be obtained to simulate the functionalities of the structural suffix tree (a generalization of the p-suffix tree) of Shibuya~\cite{Shibuya00}.
A succinct representation of the parameterized dictionary matching index of Idury and Sch{\"{a}}ffer~\cite{IduryS94} is also achieved.

\subsection{Previous and Related Work}

Parameterized pattern matching has seen constant development since its inception by Baker~\cite{Baker93} in 1993.
The $\O(n\sigma_p + n \log \sigma)$ time algorithm presented in~\cite{Baker93} for constructing p-suffix trees was improved to $\O(n \log \sigma)$ by Kosaraju~\cite{Kosaraju95}.
Later, Cole and Hariharan~\cite{ColeH03a} gave an $\O(n)$ time randomized algorithm. 
Amir, Farach, and Muthukrishnan~\cite{AmirFM94} presented an algorithm when the text $\T$ and pattern $P$ are both provided at query time. 
More recently, Jalsenius, Porat and Sach presented a solution for p-matching in the streaming model~\cite{JalseniusPS13}.
Other variants include the two-dimensional p-matching problem 
\cite{AmirACLP03}, approximate p-matching 
\cite{HazayLS04}, parameterized longest common subsequence 
\cite{KellerKL09}, and p-matching on non-linear structures 
\cite{AmirN09}. 
We refer the reader to~\cite{Lewenstein15a,MendivelsoP15} for recent surveys.

\subsection{Map} 

In Section~\ref{sec:prelim}, we first revisit a few standard text indexing data structures; in particular, we  take a close look at the parameterized suffix tree of Baker~\cite{Baker93} as it plays a crucial role in the proposed index. 
In Section~\ref{sec:pBWT}, we introduce the parameterized BWT.
In Section~\ref{sec:lfMapping}, we implement an analogous version of the last-to-first column mapping (LF mapping), introduced in the FM-index of Ferragina and Manzini~\cite{FerraginaM00}. 
We also present the key components of the proposed index in this section. 
Using the LF mapping implementation and its accompanying data structures, an adaptation of the backward search technique of Ferragina and Manzini~\cite{FerraginaM00} enables us to arrive at Theorem~\ref{thm:param:a}.
Section~\ref{sec:backwardSearch} presents the details.
We prove Theorem~\ref{thm:param:b} in Section~\ref{sec:succinctSpace}.
Section~\ref{sec:struct} is devoted to obtain a succinct representation of the structural suffix tree of Shibuya~\cite{Shibuya00}.
Section~\ref{sec:dictionary} presents a succinct representation of the parameterized dictionary matching index of Idury and Sch{\"{a}}ffer~\cite{IduryS94}.
Finally, by presenting a few unaddressed problems, we conclude the paper in Section~\ref{sec:conclusion}.

\section{Background}
\label{sec:prelim}

Throughout this paper, we use the following terminologies: for a string $S$, $|S|$ is its length, $S[i], 1\leq i \leq |S|$, is its $i$th character and $S[i,j]$ is its substring from $i$ to $j$ (both inclusive). 
If $i>j$, $S[i,j]$ denotes an empty string.
Also $S_i$ denotes the circular suffix starting at position $i$. 
Specifically, $S_i$ is $S$ if $i=1$ and is $S[i,|S|] \circ S[1,i-1]$ otherwise, where $\circ$ denotes the \emph{concatenation}. 

\subsection{Suffix Tree and Suffix Array}
\begin{figure}[!t]
\begin{center}
\includegraphics[scale=0.9]{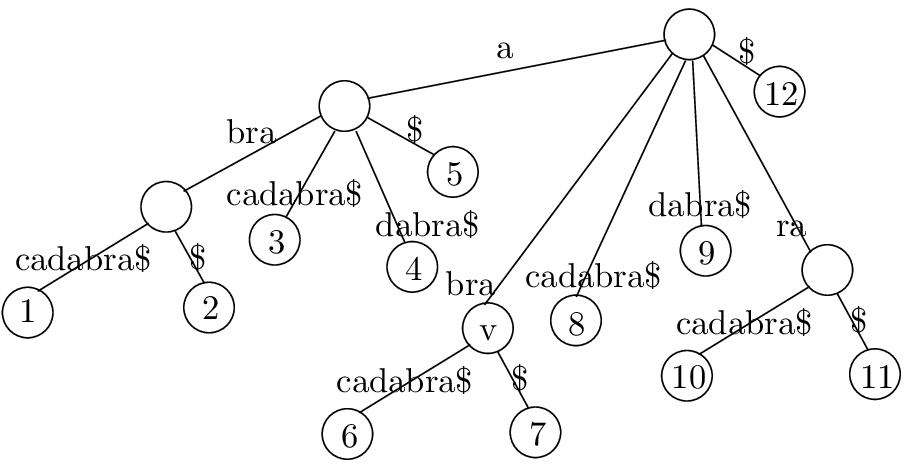}
\end{center}

\centering
\scriptsize
\begin{tabular}{|c|c|c|c|c|c|c|}
\hline
$i$ & $\T_i$ & $\SA[i]$ & $\SAI[i]$ & $\T_{\SA[i]}$ & $\BWT[i] = L[i]$ & $\LF(i) = \SAI[\SA[i]-1]$ \\ \hline
1 & abracadabra\$ & 1 & 1 & abracadabra\$ & \$ & 12 \\ \hline
2 & bracadabra\$a & 8 & 6 & abra\$abracad & d & 9 \\ \hline
3 & racadabra\$ab & 4 & 10 & acadabra\$abr & r & 10 \\ \hline
4 & acadabra\$abr & 6 & 3 & adabra\$abrac & c & 8 \\ \hline
5 & cadabra\$abra & 11 & 8 & a\$abracadabr & r & 11 \\ \hline
6 & adabra\$abrac & 2 & 4 & bracadabra\$a & a & 1 \\ \hline
7 & dabra\$abraca & 9 & 9 & bra\$abracada & a & 2 \\ \hline
8 & abra\$abracad & 5 & 2 & cadabra\$abra & a & 3 \\ \hline
9 & bra\$abracada & 7 & 7 & dabra\$abraca & a & 4 \\ \hline
10 & ra\$abracadab & 3 & 11 & racadabra\$ab & b & 6 \\ \hline
11 & a\$abracadabr & 10 & 5 & ra\$abracadab & b & 7 \\ \hline
12 & \$abracadabra & 12 & 12 & \$abracadabra  & a & 5 \\ \hline
\end{tabular}

\caption{The text is $\T[1,12] = abracadabra\$$. 
We assume the lexicographic order is $a \prec b \prec c \prec d \prec r \prec \$$.
Here, locus of $br$ is the node $v$ and suffix range of $br$ is $[6,7]$.
}
\label{fig:ST}
\end{figure}

\noindent
A suffix tree $\ST$ is a compact trie that stores all the (non-empty) suffixes of $\T$~\cite{Ukkonen:STC,Weiner:LPM}. 
Leaves in the suffix tree are numbered in the lexicographic order of the suffix they represent.
The locus of a pattern $P$ is the highest node $u$ such that $P$ is a prefix of the string formed by concatenating the edge labels from the root to $u$.
The suffix range of $P$ is the set of leaves in the subtree of $\ST$ rooted at the locus of $P$. 
See Figure~\ref{fig:ST} for an illustration.
Using suffix trees, under the perfect hashing assumption, the locus node (or equivalently, the suffix range) of any pattern $P$ can be computed in time $O(|P|)$.
Without hashing, the suffix range can be computed in time $O(|P|\log \sigma)$ by maintaining the outgoing edges in sorted order of the first character on the edge.

A suffix array $\SA$ is an array of length $n$ that maintains the lexicographic arrangement of all the suffixes of $\T$~\cite{Manber:SA}.
More specifically, if the $i$th smallest suffix of $\T$ starts at $j$, then $\SA[i] = j$ and $\SAI[j] = i$.
The former is referred to as the \emph{suffix array value} and the latter as the \emph{inverse suffix array value}.
See Figure~\ref{fig:ST} for an illustration.
Once the suffix range $[\sp,\ep]$ of $P$ is found, by using the suffix array, we can find the set $\{\SA[i] \mid \sp \leq i \leq \ep\}$ of all $occ$ occurrences of $P$.
The time needed is $O(|P| + occ)$.

\subsection{Burrows-Wheeler Transform and FM-Index}
\label{app:BWT}

Burrows and Wheeler~\cite{Burrows94} introduced a reversible transformation of the text, known as the Burrows-Wheeler Transform (BWT).
Recall that $\T_x$ is the circular suffix starting at position $x$.
Then, BWT of a text $\T$ is obtained as follows: first create a conceptual matrix $M$, such that each row of $M$ corresponds to a unique circular suffix, and then lexicographically sort all rows.  
The BWT of the text $\T$ is the last column $L$ of $M$. 
See Figure~\ref{fig:ST} for an illustration.
Note that BWT is essentially a permutation of the text $\T$. 
%
%
%
%

\subsubsection{Last-to-First Column Mapping}
The underlying principle that enables pattern matching using an FM-Index~\cite{FerraginaM00} is the last-to-first column mapping.
For any $i \in [1,n]$, $\LF(i)$ is the row $j$ in the matrix $M$ where $\BWT[i]$ appears as the first character in $\T_{\SA[j]}$.
Specifically, $\LF(i) = \SAI[\SA[i]-1]$.
Once the BWT of the text is obtained, $\LF(i)$ for any suffix $i$ is computed by the following formula:
$$\LF(i) = \CountL[\BWT[i]] + \CountE[\BWT[i],i]$$
where $\CountL[c]$ is the total number of characters in $\T$ smaller than $c$, and $\CountE[c, j]$ is the number of occurrences of $c$ in $\BWT[1,j]$.
By maintaining a Wavelet Tree~\cite{Grossi:2003:HET} over $\BWT[1,n]$, both $\CountL[\cdot]$ and $\CountE[\cdot,\cdot]$ operations can be carried out in $\O(\log \sigma)$ time.
See Section~\ref{sec:WT} for more details on the Wavelet Tree data structure.

\subsubsection{Backward Search}
Ferragina and Manzini~\cite{FerraginaM00} showed that using LF-mapping, the suffix range $[\sp,\ep]$ of a pattern $P[1,|p|]$ can be found by reading $P$ starting from the last character.
Specifically, for $i>1$, suppose the suffix range of $P[i,p]$ is known. 
Then, the suffix range of $P[i-1,p]$ can be obtained using LF-mapping.
The backward search procedure is summarized in Algorithm~\ref{alg:backwardOriginal}.
%
 \begin{algorithm}[!htb]
 \caption{for finding Suffix Range of $P[1,p]$}
 \label{alg:backwardOriginal}
 \begin{algorithmic}[1]
 \State $c \gets P[p]$, $i \gets p$
 \State $\sp \gets 1 + \CountL[c]$, $\ep \gets \CountL[c+1]$
 \While{($\sp \leq \ep$ and $i \geq 2$)}
 	\State $c \gets P[i-1]$
 	\State $\sp \gets 1 + \CountL[c] + \CountE[c,\sp-1]$
 	\State $\ep \gets \CountL[c] + \CountE[c,\ep]$
\EndWhile
 \State \textbf{if} $(\sp < \ep)$ \textbf{then} ``no match found'' \textbf{else return} $[\sp,\ep]$
 \end{algorithmic}
 \end{algorithm}
 \\
By maintaining a sampled-suffix array, which occupies $o(n)$ bits in total, each occurrence can be reported in $\O(\log ^{1 + \epsilon} n \log \sigma)$ time, where $\epsilon > 0$ is an arbitrarily small constant.
The key idea is to explicitly store $\SA$ for some specific values.
Then, a particular $\SA[i]$ can be obtained directly if the value has been explicitly stored; otherwise, it can be obtained via at most $\lceil \log ^{1 + \epsilon} n \rceil$ successive $\LF(\cdot)$ operations.

\subsection{Baker's Parameterized  Suffix Tree}
Baker~\cite{Baker93} introduced the following encoding scheme for matching strings over a parameterized alphabet $\Sigma = \Sigma_s \cup \Sigma_p$. 
A string $S$ is encoded into a string $\prev(S)$ of length $|S|$ by replacing the first occurrence of every p-character in $S$ by $0$ and any other occurrence of a p-character by the difference in text position from its previous occurrence.
More specifically, for any $i \in [1,|S|]$, $\prev(S)[i] = S[i]$ if $S[i]$ is an s-character; otherwise, $\prev(S)[i] = (i-j)$, where $j < i$ is the last occurrence (if any) of $S[i]$ before $i$. 
If $j$ does not exist, then $j = i$.
For example, $\prev(AxByCx)=A0B0C4$, where $A,B\in \Sigma_s$ and $x,y\in \Sigma_p$.  
Note that $\prev(S)$ is a string over $\Sigma'= \Sigma_s \cup \{0,1,\dots, |S|-1\}$, and can be computed in time $\O(|S|\log \sigma)$\footnote{\label{footnote:prev}
Assume that $\Sigma$ comprises of integers from $1$ through $\sigma_p$ representing the p-characters and $\sigma_p+1$ through $\sigma$ representing the s-characters.
Maintain an array $F[1,\sigma_p]$, all values initialized to $0$. 
Scan $S$ from left to right, and suppose a p-character $k \leq \sigma_p$ appears at position $x$. 
Then $\prev(S)[x]$ equals $0$ if $F[k] = 0$, and equals $x-F[k]$, otherwise.
Update $F[k] = x$.
Time is $\O(|S|)$.
For a general $\Sigma$, we take the convention that all p-characters are lexicographically smaller than s-characters.
Then, the string $S$ over $\Sigma$ can be converted into a string over an integer alphabet (as described previously) using a balanced binary search tree in $\O(|S|\log \sigma)$ time by simply mapping the $k$th smallest symbol to the integer $k$.
}.

\BF[\cite{Baker93}]
\label{fact:bakerPrev}
Two strings $S$ and $S'$ are a p-match iff $\prev(S) = \prev(S')$.
Also $S$ and a prefix of $S'$ are a p-match iff $\prev(S)$ is a prefix of $\prev(S')$.
\EF
Moving forward, we follow the  convention below. 
\BC
\label{convention:lexico}
In $\Sigma'$, the integer characters (corresponding to p-characters) are lexicographically smaller than s-characters\footnote{
If $\Sigma_s$ contains integers, then we simply map the $k$th smallest s-character to the integer $|S|+k$ to ensure that it is disjoint from the set $\{0,1,\dots,|S|-1\}$.
The string $|S|$ can be pre-processed in $O(|S|\log \sigma)$ time using a balanced binary search tree to ensure that this condition holds.
This pre-processing will not affect any time complexities.}. 
An integer character $i$ comes before another integer character $j$ iff $i < j$. 
Also, $\$$ is lexicographically larger than all other characters. 
\EC
Parameterized Suffix Tree ($\pST$) is the compacted trie of all strings in $\calP = \{\prev(\T[k,n]) \mid 1\leq k \leq n\}$. 
Each edge is labeled with a string over $\Sigma'$. 
We use $\nodePath(u)$ to denote the concatenation of edge labels on the path from root to node $u$. 
Clearly, $\pST$ consists of $n$ leaves (one per each encoded suffix) and at most $n-1$ internal nodes. 
The space required is $\Theta(n\log n)$ bits.
See Figure~\ref{fig:pSAandpBWT} for an illustration. 
The path of each leaf node corresponds to the encoding of a unique suffix of $\T$, and leaves are ordered in the lexicographic order of the corresponding encoded suffix.
 
To find all occurrences of $P$, traverse $\pST$ from root by following the edges labels and find the highest node $u$ (called \emph{locus}) such that $\nodePath(u)$ is prefixed by $\prev(P)$. 
Then find the range $[\sp,\ep]$ (called \emph{suffix range} of $\prev(P)$) of leaves in the subtree of $u$ and report  $\{\pSA[i] \mid \sp \leq i \leq \ep\}$ as the output. 
Here, $\pSA[1,n]$ is the parameterized suffix array i.e., $\pSA[i] = j$ and $\pSAI[j] = i$ iff $\prev(\T[j,n])$ is the $i$th lexicographically smallest string in $\mathcal{P}$.
Note that $\nodePath(\ell_i)=\prev(\T[\pSA[i],n])$, where $\ell_i$ is the $i$th leftmost leaf in $\pST$. 
The query time is $\O(|P|\log \sigma +occ)$
\footnote{
The query time is achieved by using perfect hashing, i.e., at each node, we can navigate to its appropriate child in constant time.
The $O(|P| \log \sigma)$ factor is due to the time to obtain $\prev(P)$.
If we remove perfect hashing, then the time increases to $\O(|P|\log n +occ)$, as each internal node may have $\Theta(n)$ children.
}.

\section{Parameterized Burrows-Wheeler Transform}
\label{sec:pBWT}

We introduce a similar transform to that of the BWT, which we call the Parameterized Burrows-Wheeler Transform (p-BWT).
To obtain the p-BWT of  $\T$, we create a similar matrix $M$ as in the case of the BWT described earlier. 
Then, we sort all this rows lexicographically according to the $\prev(\cdot)$ encoding of the corresponding unique circular suffix,  and obtain the last column $L$ of the sorted matrix $M$. 
Clearly, the $i$th row is equal to $\T_{\pSA[i]}$. 
Moving forward, denote by $f_i$, the first occurrence of $L[i]$ in $\T_{\pSA[i]}$.
The p-BWT of $\T$, denoted by $\pBWT[1,n]$, is defined as follows: 
\begin{equation*}
\pBWT[i] =
\begin{cases}
L[i], & \text{if } L[i] \text{ is an s-character},
\\
\text{number of distinct p-characters in }\T_{\pSA[i]}[1,f_i],&  \text{otherwise}. 
\end{cases}
\end{equation*}

\begin{figure}[t]

\begin{center}
\includegraphics[scale=0.8]{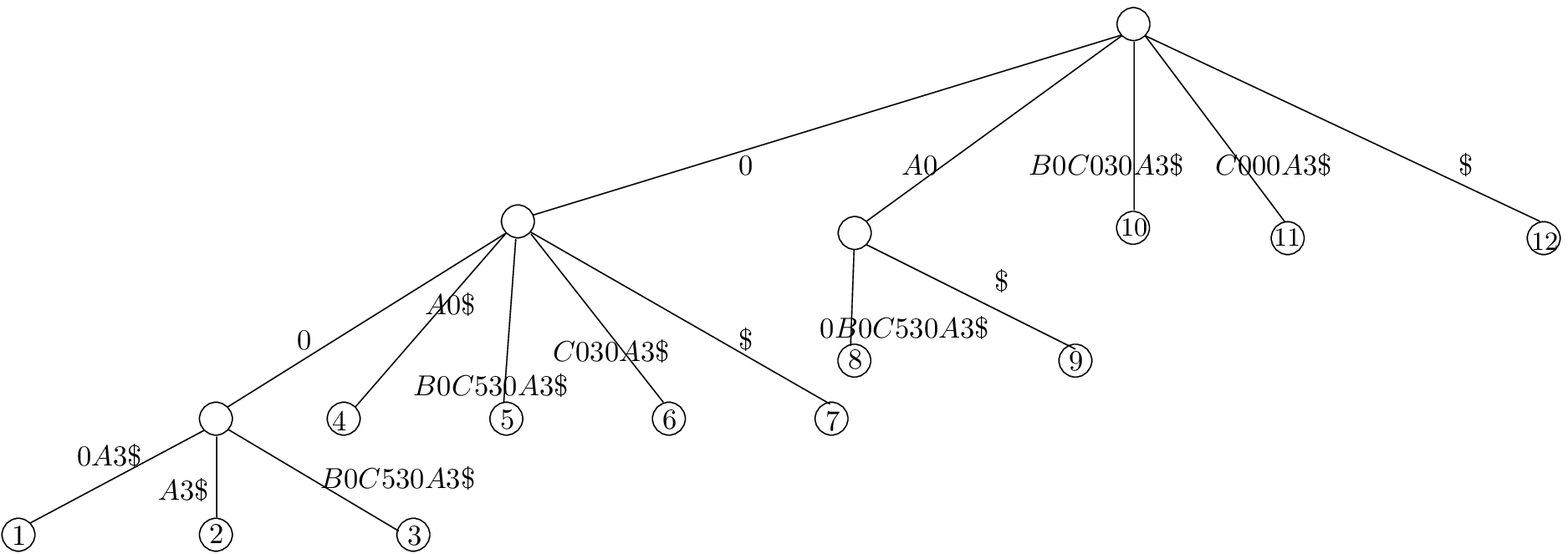}
\end{center}

\scriptsize
\centering
\begin{tabular}{|c|c|c|c|c|c|c|c|c|c|}
\hline
$i$ & $\T_i$ & $\prev(\T_i)$ &  $\prev(\T_{\pSA[i]})$ & \begin{tabular}[c]{@{}c@{}}  $\T_{\pSA[i]}$\end{tabular} & $\pSA[i]$ & $L[i]$ & $f_i$ & $\pBWT[i]$ & $\pLF(i)$ \\ \hline
1 & AxyBzCxzwAz\$ & A00B0C530A3\$ & 000A3\$A70B6C & xzwAz\$AxyBzC & 7 & C &   & C & 11\\ 
2 & xyBzCxzwAz\$A & 00B0C530A3\$A & 00A3\$A00B6C5 & zwAz\$AxyBzCx & 8 & x & 7 & 3 & 1\\
3 & yBzCxzwAz\$Ax & 0B0C030A3\$A7 & 00B0C530A3\$A & xyBzCxzwAz\$A & 2 & A &   & A & 8\\
4 & BzCxzwAz\$Axy & B0C030A3\$A70 & 0A0\$A00B6C53 & wAz\$AxyBzCxz & 9 & z & 3 & 2 & 2\\
5 & zCxzwAz\$AxyB & 0C030A3\$A70B & 0B0C030A3\$A7 & yBzCxzwAz\$Ax & 3 & x & 5 & 3 & 3\\
6 & CxzwAz\$AxyBz & C000A3\$A70B6 & 0C030A3\$A70B & zCxzwAz\$AxyB & 5 & B &   & B & 10\\
7 & xzwAz\$AxyBzC & 000A3\$A70B6C & 0\$A00B5C530A & z\$AxyBzCxzwA & 11 & A &  & A & 9\\
8 & zwAz\$AxyBzCx & 00A3\$A00B6C5 & A00B0C530A3\$ & AxyBzCxzwAz\$ & 1 & \$ &  & \$ & 12\\
9 & wAz\$AxyBzCxz & 0A0\$A00B6C53 & A0\$A00B5C530 & Az\$AxyBzCxzw & 10 & w & 12 & 4 & 4\\
10 & Az\$AxyBzCxzw & A0\$A00B5C530 & B0C030A3\$A70 & BzCxzwAz\$Axy & 4 & y & 12 & 4 & 5\\
11 & z\$AxyBzCxzwA & 0\$A00B5C530A & C000A3\$A70B6 & CxzwAz\$AxyBz & 6 & z & 3 & 2 & 6\\
12 & \$AxyBzCxzwAz & \$A00B0C530A3 & \$A00B0C530A3 & \$AxyBzCxzwAz & 12 & z & 6 & 3 & 7\\
\hline
\end{tabular}
\caption{The text is $\T[1,12] = AxyBzCxzwAz\$$, where $\Sigma_s = \{A,B,C,\$\}$ and $\Sigma_p=\{w,x,y,z\}$}
\label{fig:pSAandpBWT}
\end{figure}

In other words, when $L[i] \in \Sigma_s$,
$\pBWT[i]= \T[\pSA[i]-1]$ (define $\T[0]=\T[n]=\$$) and when $L[i] \in \Sigma_p$, 
$\pBWT[i]$ is the number of $0$'s in the $f_i$-long prefix of $\prev(\T_{\pSA[i]})$.
Thus, $\pBWT$ is a sequence of $n$ characters over the set $\Sigma'' = \Sigma_s \cup \{1,2,\dots, \sigma_p\}$ of size $  \sigma_s+\sigma_p=\sigma$.
See Figure~\ref{fig:pSAandpBWT} for an illustration of $\pSA$ and $\pBWT$.

In order to represent $\pBWT$ in succinct space, we map each s-character in $\Sigma''$ to a unique integer in $[\sigma_p+1, \sigma]$. 
Specifically, the $i$th smallest s-character will be denoted by $(i+\sigma_p)$. 
Moving forward, $\pBWT[i] \in [1,\sigma_p]$ iff $L[i]$ is a p-character and $\pBWT[i] \in [\sigma_p+1,\sigma]$ iff $L[i]$ is a s-character.
We summarize the relation between $\prev(\T_{\pSA[i]})$ and $\prev(\T_{\pSA[i]-1})$ below. 
\BO 
\label{obs:prevOcc} 
For any $1\leq i \leq n$, 
\begin{equation*}
\prev(\T_{\pSA[i]-1})=
\begin{cases}
\pBWT[i] \circ \prev(\T_{\pSA[i]})[1,n-1],
 & \text{if } \pBWT[i] \in \Sigma_s,\\
0 \circ \prev(\T_{\pSA[i]})[1,f_i-1] \circ f_i \circ \prev(\T_{\pSA[i]})[f_i+1,n-1], &  \text{otherwise}. 
\end{cases}
\end{equation*}
\EO

\subsection{Parameterized Last-to-First Column Mapping}

Based on the conceptual matrix $M$, the {parameterized last-to-first column $(\pLF)$ mapping} of $i$ is the position at which the character at $L[i]$  lies in the first column of $M$. 
Specifically, $\pLF(i)=\pSA^{-1}[\pSA[i]-1]$. 
The significance of  $\pLF$ mapping is summarized in Lemma~\ref{lem:saiSA}. 
%
\BL 
\label{lem:saiSA}
Assume $\pLF(\cdot)$ can be computed in $t_{\pLF}$ time. 
Then, for any parameter $\Delta$, by using an additional $\O((n/\Delta)\log n)$-bit structure, we can compute $\pSA[\cdot]$ and $\pSA^{-1}[\cdot]$ in $\O(\Delta \cdot t_{\pLF})$ time. 
\EL
\begin{proof}
Define, $\pLF^0(i)=i$ and $\pLF^k(i)=\pLF(\pLF^{k-1}(i))=\pSA^{-1}[\pSA[i]-k]$ for any integer $k>0$. 
We maintain two $\Delta$-sampled arrays, one each for $\pSA$ and $\pSAI$. 
More specifically, we explicitly maintain $\pSA[j]$ and $\pSAI[j]$ if the value belongs to $\{1, 1+\Delta, 1+2\Delta, 1+3\Delta, \dots, n\}$. 
The total space for each sampled array can be bounded by $\O((n/\Delta)\log n)$ bits.
To find $\pSA[i]$,  repeatedly apply the $\pLF(\cdot)$ operation (starting from $i$) until you obtain a $j$ such that $\pSA[j]$ has been explicitly stored.
Suppose, the number of such operations is $k$. 
Then, $j = \pLF^k(i) = \pSA^{-1}[\pSA[i]-k]$, which gives $\pSA[i] = \pSA[j] + k$.
Since $k \leq \Delta$, $\pSA[i]$ is computed in $\O(\Delta \cdot t_{\pLF})$ time.
To find $\pSAI[i]$, find the smallest $j \geq i$ whose $\pSAI[j]$ is explicitly stored. 
Then, $\pSAI[i] = \pLF^{j-i}(\pSAI[j])$. 
As $j-i \leq \Delta$, the time is bounded by $\O(\Delta \cdot t_{\pLF})$.
\qed
\end{proof}
\paragraph{•}
We have the following as a consequence of Lemma~\ref{lem:saiSA}.
\BCR
\label{coro:extract}
Assume $\pLF(\cdot)$ can be computed in $t_{\pLF}$ time and $\pBWT[\cdot]$ can be accessed in $t_{\pBWT}$ time. 
Then, for any parameter $\Delta$, using an additional $\O((n/\Delta)\log n)$-bit structure, $\prev(\T[x,y])$ can be extracted in $\O(\Delta\cdot t_{\pLF} +  d(t_{\pLF}+t_{\pBWT}+\log \sigma))$ time, where $d = y-x+1$. 
\ECR
\begin{proof}
Note that $\T[n] = \$$ and by Convention~\ref{convention:lexico}, we have $\pSA[n] = \pSAI[n] = n$.
We first retrieve the lexicographic position $i$ of the circular suffix $\T_{y+1}$ by using $\pSAI[y+1]$, and then retrieve $\pBWT[i]$.
This is achieved in $\O(\Delta \cdot t_{\pLF} + t_{\pBWT})$ time.
We now proceed to retrieve the character $\pBWT[\pSAI[y]]$. 
Note that although this can be directly achieved using $\pSAI$, this will cost $\O(\Delta \cdot t_{\pLF} + t_{\pBWT})$ time. 
To achieve this in a more efficient way, we first find the position $i'$, such that the first character in suffix $\T_{\pSA[i']}$ is exactly the same character as $L[i]$.
Therefore, $i' = \pLF(i)$ and we obtain $\pBWT[i']$.
The time required for this step is $\O(t_{\pLF} + t_{\pBWT})$.
We repeat this process until we obtain the lexicographic position $i''$ of the circular suffix $\T_{x+1}$, and terminate by obtaining $\pBWT[i'']$.
At this point, we have obtained the following string 
$$S=\pBWT[\pLF^{d}(i)] \circ \pBWT[\pLF^{d-1}(i)] \circ \dots \circ \pBWT[\pLF^2(i)] \circ \pBWT[\pLF(i)] \circ \pBWT[i]$$
where $i = \pSAI[y+1]$. 
The time required to obtain the string is $\O(\Delta\cdot t_{\pLF} +  d(t_{\pLF}+t_{\pBWT}))$.

Recall that $\pBWT[j]$ is $L[j]$ if $L[j]$ is an s-character; otherwise, it is number of distinct p-characters in $\T_{\pSA[j]}[1,f_{j}]$, where $f_{j}$ is the first occurrence of $L[j]$ in $\T_{\pSA[j]}$.
Therefore, the obtained string contains the desired s-characters of $\T[x,y]$ at the correct position, and we are required to obtain the encoding corresponding to the  p-characters in $\T[x,y]$. 

We scan the obtained string $S$ from right to left. 
Suppose, we encounter a p-character in $S$ at a position $d'$ from left (which can be checked by comparing the $S[d']$ with $\sigma_p$). 
We find the number of distinct p-characters, say $k'$, in $S[d'+1,d]$.
This can be achieved in constant time\footnote{
We maintain a bit-vector $B[1,\sigma_p]$, all values initialized to $0$, and a counter $C$ initialized to $0$.
Scan the text from right to left, and whenever a p-character $c_p$ is encountered at position $x$, the required count at $x$ is $C$.
If $B[c_p] = 0$, then increment $C$ by $1$, set $B[c_p]$ to $1$, and continue.
}.
(In case, $d' = d$, then $k' = 0$.)
Observe that if $S[d'] > k'$, then the p-character $\T[x-1+d']$ does not appear in $\T[x+d',y]$, and we replace $S[d']$ by a new p-character $c_{k'+1}$. 
Also, insert $c_{k'+1}$ into a \emph{balanced binary search tree} keyed by $d'$.
Otherwise, $\T[x-1+d']$ is the $S[d']$th smallest p-character in the tree. 
We retrieve this character from the tree and update the key to $d'$. 
Since, the tree has at most $\sigma_p$ nodes, insertion, search, and key update operations are performed in $\O(\log \sigma_p)$ time.
Therefore, this process takes $\O(d \log \sigma_p)$ time.
Finally, we encode the string using $\prev(\cdot)$ in $\O(d\log \sigma_p)$ time, which results in the claimed complexity.
\qed
\end{proof}

\paragraph{•}
To aid the reader's intuition for computing $\pLF$ mapping, we present Lemma~\ref{lem:atleastOneStatic}, which shows how to compare the lexicographic rank of two encoded suffixes when prepended by their respective previous characters.
This key concept is implemented in Sections~\ref{sec:lfMapping} and~\ref{sec:succinctSpace} to arrive at Theorem~\ref{thm:lfmapping}.
\BL 
\label{lem:atleastOneStatic}
Consider two suffixes $i$ and $j$ corresponding to the leaves $\ell_i$ and $\ell_j$ in $\pST$.
Then, $\pLF(i)$ and $\pLF(j)$ are related as follows:
\begin{enumerate}[label={\emph{(\alph*)}}]
\item 
If $L[i]$ is parameterized and $L[j]$ is static, then $\pLF(i) <\pLF(j)$.
\item 
If both $L[i]$ and $L[j]$ are static, then $\pLF(i) < \pLF(j)$ iff one of the following holds: 
\begin{itemize}
\item $\pBWT[i] < \pBWT[j]$
\item $\pBWT[i] = \pBWT[j]$ and $i<j$
\end{itemize}
\item 
Assume that both $L[i]$ and $L[j]$ are parameterized and $i<j$. 
Let $u$ be the lowest common ancestor of $\ell_i$ and $\ell_j$ in $\pST$, and $z$ be the number of $0$'s in the string $\nodePath(u)$.
Then,
\begin{enumerate}[label={\emph{(\arabic*)}}]
\item
If $\pBWT[i], \pBWT[j] \leq z$, then $\pLF(i) < \pLF(j) \text{ iff } \pBWT[i] \geq \pBWT[j]$.
\item
If $\pBWT[i]\leq z <\pBWT[j]$, then $\pLF(i) > \pLF(j)$.
\item
If $\pBWT[i]> z \geq \pBWT[j]$, then $\pLF(i) < \pLF(j)$.
\item
If $\pBWT[i], \pBWT[j] > z$, then $\pLF(i) >\pLF(j)$ iff all of the following are true: 
\begin{itemize}
\item $\pBWT[i] = z+1$.
\item the leading character on the $u$ to $\ell_i$ path is $0$.
\item the leading character on the $u$ to $\ell_j$ path is not an s-character.
\end{itemize}
\end{enumerate}
\end{enumerate}
\EL 
\begin{proof}
(a) and (b): Follows immediately from Convention~\ref{convention:lexico} and Observation~\ref{obs:prevOcc}.\\
(c) Recall that $f_i$ and $f_j$ are the first occurrences of the characters $L[i]$ and $L[j]$ in the circular suffixes $\T_{\pSA[i]}$ and $\T_{\pSA[j]}$ respectively.
Let $d = |\nodePath(u)|$. 
Clearly, the conditions $(1)$--$(4)$ can be written as:
(1) Both $f_i,f_j \leq d$, (2) $f_i \leq d$ and $f_j > d$, (3) $f_i > d$ and $f_j \leq d$, and (4) Both $f_i,f_j > d$.

Then the claims $(1)$--$(3)$ are immediate from Observation~\ref{obs:prevOcc} and Convention~\ref{convention:lexico}.
For proving $(4)$, first observe that if $\T_{\pSA[j]}[d+1]$ is an s-character, then $\T_{\pSA[j]-1}[d+2] > \T_{\pSA[i]-1}[d+2]$, and $\pLF(i) < \pLF(j)$.
So, assume otherwise.
Let $e_i$ and $e_j$ be the $(d+1)$th characters of $\prev(\T_{\pSA[i]})$ and $\prev(\T_{\pSA[j]})$ respectively.
Since, the suffixes $i$ and $j$ separates after $u$, $f_i \neq f_j$.
Also, $i<j$ implies $0 \leq e_i < e_j \leq d$.
Note that if $\pBWT[i] = z+1$ and $e_i=0$, then $L[i] = \T_{\pSA[i]}[d+1]$ i.e., $f_i = d+1$, and $\prev(\T_{\pSA[i]-1})[d+2] = d+1 > e_j = \prev(\T_{\pSA[j]-1})[d+2]$.
Otherwise, $\prev(\T_{\pSA[i]-1})[d+2] = e_i < e_j \leq \prev(\T_{\pSA[j]-1})[d+2]$.
\qed
\end{proof}

\BT
\label{thm:lfmapping}
We can compute $\pLF(i)$ as follows: 
\begin{enumerate}[label={\emph{(\alph*)}},ref={\thethm\alph*}]
\item \label{thm:lfmapping:a} in $\O(\log \sigma)$ time using $\O(n\log \sigma)$ bits.
\item \label{thm:lfmapping:b} in $\O(\log \sigma \log \log \sigma)$ time using $n\log\sigma+\O(n)$ bits.
\end{enumerate}
\ET

\section{Implementing $\pLF$ Mapping in Compact Space} 
\label{sec:lfMapping}
We prove Theorem~\ref{thm:lfmapping:a} in this section. 
\subsection{Data Structure Toolkit}
The following form the key components of our data structure. 

\subsubsection{Wavelet Tree over $\pBWT$}
\label{sec:WT}

Grossi, Gupta, and Vitter~\cite{Grossi:2003:HET} introduced the wavelet tree (WT) data structure, which generalizes the well-known $\rank$ and $\select$ queries over bit-vectors\footnote{
Given a bit-vector $B$ and $c \in \{0,1\}$, $\rank(i,c) = |\{j \mid j \leq i \text{ and } B[j] = c\}|$ and $\select(i,c) = \min \{j \mid \rank(j,c) = i\}$
}. 
Specifically, given an array $A$ over an alphabet $\Sigma$, by maintaining a data structure of size $|A|\log |\Sigma| + o(|A|\log |\Sigma|)$ bits, the following queries can be supported in $\O(\log |\Sigma|)$ time~\cite{Grossi:2003:HET,GagieNP12,WT}:
\begin{itemize}
\item $A[i]$.
\item $\rank_A(i,x)=$ the number of occurrences of $x$ in $A[1,i]$.
\item $\select_A(i,x)=$ the $i$th occurrence of $x$ in $A[1,n]$.
\item $\pred_A(i,x)=j$, the rightmost position before $i$ such that $A[j] <x$.
\item $\rangeCount_A(i,j,x,y)=$ number of elements in $A[i,j]$ that are at least $x$ and at most $y$.
\item $\rangeNV_A(i,j,x,k)=$ the $k$th smallest value in $A[i,j]$ that is greater than $x$.
\item $\rangePV_A(i,j,x,k)=$ the $k$th largest value in $A[i,j]$ that is smaller than $x$.
\end{itemize}
We drop the subscript $A$ when the context is clear.
The $\pBWT$ is a string of length $n$ over an alphabet set $\Sigma'' =\Sigma_s \cup \{1,2,\dots, \sigma_p\}$ of size $\sigma = \sigma_s + \sigma_p$. 
By maintaining a WT over $\pBWT$ in $n\log\sigma+o(n\log\sigma)$ bits, we can support the above operations over the $\pBWT$.
As noted by Navarro~\cite{WT}, we can apply the technique of Golynski et al.~\cite{GolynskiGGRR07} to reduce the redundancy of $o(n\log \sigma)$ bits to $o(n)$ bits. 
The time to answer any of the above queries remains unaffected. 

\subsubsection{Succinct representation of the topology of $\pST$}
\label{sec:treeTopology}

We rely on the following result of Navarro and Sadakane~\cite{NavarroS14,SadakaneN10}. 
Any tree having $m$ nodes can be represented in $2m+o(m)$ bits, such that if each node is labeled by its pre-order rank in the tree, the following operations can be supported in constant time (note that $m<2n$ in our case):
\begin{itemize}

\item $\pre(u)$ = the pre-order rank of node $u$.
\item $\parent(u)$ = the parent of node $u$.
\item $\nd(u)$ = the number of edges on the path from root to $u$.
\item $\child(u,q)$ = the $q$th leftmost child of node $u$.
\item $\lca(u,v)$ = the lowest common ancestor (LCA) of two nodes $u$ and $v$.
\item $\lmostLeaf(u)$/$\rmostLeaf(u)$ = the leftmost/rightmost leaf of the subtree rooted at $u$.
\item $\la(u,D)$ = the ancestor of $u$ such that $\nd(u) = D$. 

\end{itemize}
The representation also allows us to find the pre-order rank of the $i$th leftmost leaf in $\O(1)$ time.
Thus, for a particular suffix $i$, we can locate its corresponding leaf $\ell_i$ in this representation (i.e., find the pre-order rank of $\ell_i$) in $\O(1)$ time.
Given a leaf $\ell$,  in $\O(1)$ time, we can also find the number of suffixes which appear before the suffix corresponding to $\ell$.
Moving forward, we will use $\ell_i$ to denote the leaf corresponding to the $i$th lexicographically smallest $\prev$-encoded suffix.

\subsubsection{Marked and Prime Nodes}
\label{sec:marked}

We use the following scheme of Hon et al.~\cite{HonSTV14}.
Identify certain nodes in a tree ($\pST$ in our case) as \emph{marked nodes}. 
Let $g$ be a parameter called grouping factor. 
Starting from the leftmost leaf in $\pST$, we combine every $g$ leaves together to form a group. 
In particular, the leaves $\ell_1$ through $\ell_g$ form the first group, $\ell_{g+1}$ through $\ell_{2g}$ form the second, and so on. 
We mark the LCA of the first and last leaves of every group. 
Moreover, for any two consecutive marked nodes in pre-order, we mark their LCA as well, 
and continue recursively.
Note that the root node is marked. 
For every marked node $u^*$, we also identify a \emph{prime node}, which is the child of the lowest marked ancestor $u'$ of $u^*$ that lies on the path from $u'$ to $u^*$.
If $\parent(u^*)$ is marked then $u^*$'s prime node is $u^*$ itself.
\BF[\cite{HonSTV14}]
\label{fact:marked}
The following properties are ensured by the above marking scheme. 
\begin{enumerate}[label={\emph{(\arabic*)}}]

\item The number of marked nodes and prime nodes is $\O(n/g)$. 

\item For any (prime) node $u'$, the highest marked node $u^*$ in its subtree is unique. 
Let $\Leaf(v)$ be the set of leaves in the subtree of a node $v$.
Then, $\Leaf(u')\setminus \Leaf(u^*)$ is of size at most $2g$. 
Also, for a lowest marked node $u$, $|\Leaf(u)| \leq 2g$.

\item The path from a (marked) node to its nearest marked ancestor contains at most $g$ nodes. 

\item Given a node, in $\O(1)$ time, we can verify whether it has a marked descendant or not. 
We can also find the highest marked descendant (if any) in $\O(1)$ time\footnote{
Let $i$ and $j$ be two multiples of $g$ such that $j > i \geq 0$.
Clearly, for any such $i,j$ pair, if the subtree of $u$ contains leaves $\ell_{i+1}$ and $\ell_j$, then there $u$ has a marked descendant, otherwise not.
The LCA of such a maximal pair of leaves $\ell_{i+1}$ and $\ell_j$ gives the highest marked descendant.
}.

\item Given a node, in $\O(1)$ time, we can find its lowest marked ancestor and its lowest prime ancestor (if any) using an  $\O(n)$-bit structure\footnote{
Maintain a bit-vector $B$ such that $B[i] = 1$ iff the node with pre-order rank $i$ is a marked node.
Additionally, maintain a $\rank$-$\select$ data structure over this bit-vector.
Now given a query node with pre-order rank $j$, using these data structures, in constant time, find the largest $j' < j$ such that $B[j'] = 1$ and the smallest $j'' > j$ such that $B[j''] = 1$.
Note that one of $j'$ or $j''$ must exist.
If $j'$ is an ancestor of $j$ (which can be verified by comparing the ranges of the leaves in their subtree), then $j'$ is the desired lowest marked ancestor.
In case, $j'$ or $j''$ does not exist, we take them to be the root node.
Consider the nodes $\lca(j,j')$, $\lca(j,j'')$, and $\lca(j',j'')$. 
The desired lowest marked ancestor $u$ is the one among these three which is marked and has maximum node depth.
Once $u$ is retrieved, we check whether it has marked descendant, or not. 
If it has one, the lowest prime ancestor is given by $\la(j,\nd(u)+1)$.
}.
\end{enumerate}
\EF

\subsection{ZeroDepth and ZeroNode}
For a node $u$, $\alphaDepth(u)$ is the number of $0$'s in $\nodePath(u)$.
For a leaf $\ell_i$ with $\pBWT[i]\in [1,\sigma_p]$, $\zeroNode(\ell_i)$ is the highest node $z$ on the root to $\ell_i$ path such that $\alphaDepth(z) \geq \pBWT[i]$. 
Thus, $z$ is the locus of $\nodePath(\ell_i)[1,f_i]$.
Note that $z$ necessarily exists as $\alphaDepth(\ell_i) \geq \pBWT[i]$.
Moving forward, whenever we refer to $\zeroNode(\ell_i)$, we assume $\pBWT[i] \in [1,\sigma_p]$.

\BL 
\label{lem:zeronode}
We can compute $\zeroNode(\ell_i)$ in $\O(\log\sigma)$ time using an 
$\O(n\log\sigma)$-bit structure.
\EL
\begin{proof}
We find the lowest node $u$ on the root to $\ell_i$ path that satisfies $\alphaDepth(u) < \pBWT[i]$. 
Clearly, $\zeroNode(\ell_i)$ is the child of $u$ on this path, and is given by $\la(\ell_i,\nd(u)+1)$.
Since the $\alphaDepth$ of the root node is $0$, $u$ necessarily exists.

Maintain an array $D$ such that $D[k]$ equals $\alphaDepth$ of the node with pre-order rank $k$. 
Instead of maintaining $D$ explicitly, we maintain a wavelet tree over it.
The space required is $2n \log \sigma + o(n)$ bits.
We locate the node, say $v$, with pre-order rank $\pred_D(j,\pBWT[i])$, where $j$ is the pre-order rank of $\ell_i$.
Then, $u$ is given by $\lca(\ell_i,v)$.
The time required is $\O(\log \sigma)$.

To see the correctness first observe that for any node $x$, $\alphaDepth(x) \geq \alphaDepth(\parent(x))$.
Thus, $\alphaDepth(\lca(\ell_i,v)) \leq \alphaDepth(v) < \pBWT[i]$.
If $\lca(\ell_i,v)$ is not the desired node, then its child $u'$ on the path to $\ell_i$ satisfies $\alphaDepth(u') < \pBWT[i]$. 
But then $u'$ appears after $v$ and before $\ell_i$ in pre-order, a contradiction.
\qed
\end{proof}

\paragraph{•}
We remark that the following additional functionalities: $\leafc(\cdot)$, $\fs(\cdot)$ and $\pc(\cdot)$ will be defined later.
Each of these can be computed in $\O(1)$ time using an $\O(n)$-bit structure. 
In summary, the total space required by the above data structures is $\O(n \log \sigma)$ bits.

\subsection{Computing $\pLF(i)$ when $\pBWT[i] \in  [\sigma_p+1, \sigma]$}

In this case, $L[i]=\pBWT[i]$ is an s-character.
Using Lemma~\ref{lem:atleastOneStatic}, we conclude that $\pLF(i) > \pLF(j)$ iff either $j \in [1,n]$ and $\pBWT[j] < \pBWT[i]$, or $j \in [1,i-1]$ and $\pBWT[i] = \pBWT[j]$.
Then, $\pLF(i) = 1+\rangeCount(1,n, 1, c-1)+\rangeCount(1,i-1,c,c)$, where $c = \pBWT[i]$.

\subsection{Computing $\pLF(i)$ when $\pBWT[i] \in  [1,\sigma_p]$}

In this case, $L[i]$ is a p-character. 
Let $z = \zeroNode(\ell_i)$ and $v = \parent(z)$.
Then, $f_i = (|\nodePath(v)|+1)$ if the leading character on the edge from $v$ to $z$ is $0$ and $\pBWT[i] = (\alphaDepth(v) + 1)$; otherwise, $f_i > (|\nodePath(v)|+1)$.
For a leaf $\ell_j$ in $\pST$, $\leafc(j)$ is a boolean variable, which is $0$ iff $f_j = (|\nodePath(\parent(\zeroNode(\ell_j)))|+1)$.
Using this information, in constant time, we can determine which of the following two cases the suffix corresponding to $\ell_i$ satisfies.
\begin{figure}[!t]
\begin{center}
	\begin{subfigure}[t]{0.45\textwidth}
    	\centering
		\includegraphics[scale=0.8]{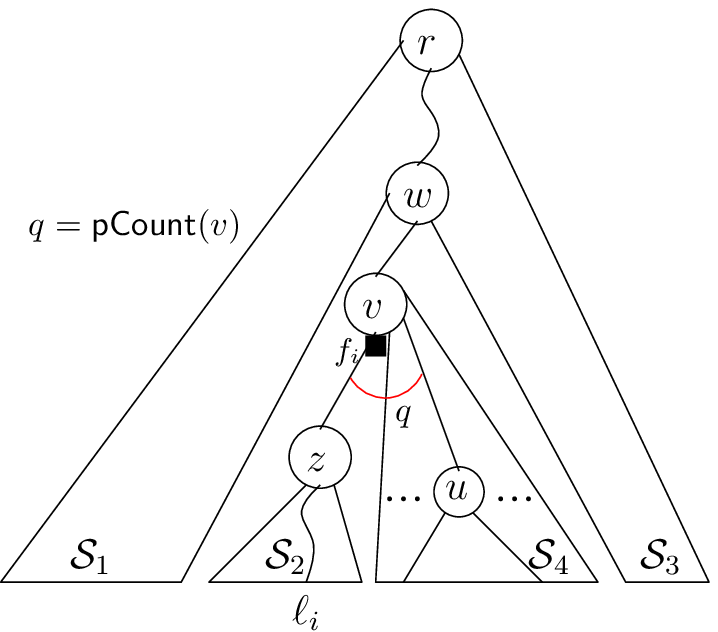}
		\caption{$f_i = |\nodePath(v)|+1$} 
		\label{fig:leadingLF}
	\end{subfigure}
	\hspace{2mm}
	\begin{subfigure}[t]{0.45\textwidth}
    	\centering
		\includegraphics[scale=0.8]{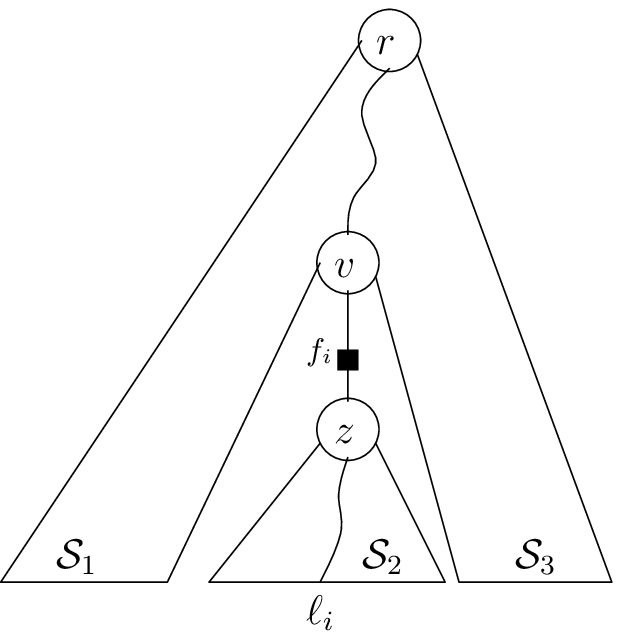}
		\caption{$f_i > |\nodePath(v)|+1$}
		\label{fig:nonLeadingLF} 
	\end{subfigure}
\end{center}
\caption{Illustration of various suffix ranges when the suffix $\T_{\pSA[i]}$ is preceded by a p-character}
\label{fig:leadNonLeadLF}
\end{figure}

\subsubsection{Case 1 ($f_i = |\nodePath(v)|+1$)} 
See Figure~\ref{fig:leadingLF} for an illustration.
Let $w$ be the parent of $v$.
Note that in this case $z$ is the leftmost child of $v$.
We partition the leaves into four sets $\s_1$, $\s_2$, $\s_3$, and $\s_4$:
\begin{itemize}

\item $\s_1$: leaves to the left of the subtree of $v$.

\item $\s_2$: leaves in the subtree of $z$.

\item $\s_3$: leaves to the right of the subtree of $v$.

\item $\s_4$: leaves in the subtree of $v$ but not of $z$.

\end{itemize}
In case, $v$ is the root node $r$, we take $w$ to be $r$ as well.
Consequently, $\s_1 = \s_3 = \varnothing$.

\subsubsection{Case 2 ($f_i > |\nodePath(v)|+1$)} 
See Figure~\ref{fig:nonLeadingLF} for an illustration.
We partition the leaves into three sets $\s_1,\s_2$, and $\s_3$:
\begin{itemize}

\item $\s_1$ (resp. $\s_3$): leaves to the left (resp. right) of the subtree of $z$.

\item $\s_2$: leaves in the subtree of $z$.

\end{itemize}

\subsubsection{Computing $\LF(i)$}
We first compute $z =\zeroNode(\ell_i)$ using Lemma~\ref{lem:zeronode}, and then locate $v = \parent(z)$. 
Using $\leafc(i)$ and the $\lmostLeaf(\cdot)$/$\rmostLeaf(\cdot)$ tree operations, we find the desired ranges.

We shall use $[L_x,R_x]$ to denote the range of leaves in the subtree of any node $x$. 
In order to compute $\pLF(i)$, we first compute $N_1$, $N_2$, and $N_3$, which are respectively the number of leaves $\ell_j$ in ranges $\s_1$, $\s_2$, and $\s_3$ such that $\pLF(j) \leq \pLF(i)$. 
Likewise, we compute $N_4$ (w.r.t $\s_4$) if we are in the first case.
Then, $\pLF(i) = N_1+N_2+N_3+N_4$.
We present the detailed description below. 

\begin{itemize}

\item \textbf{Computing $N_1$:} 
For any leaf $\ell_j \in \s_1$, $\pLF(j) < \pLF(i)$ iff $f_j > 1+|\nodePath(\lca(z,\ell_j))|$ and $L[j] \in \Sigma_p$. 
Therefore, $N_1$ is the number of leaves $\ell_j$, $L[j] \in \Sigma_p$, which comes before $z$ in pre-order with $f_j > 1+|\nodePath(\lca(z,\ell_j))|$.

Define, $\fc(x)$ of a node $x$ as the number of leaves $\ell_j$ in $x$'s subtree such that $|\nodePath(y)| + 2 \leq f_j \leq |\nodePath(x)|+1$, where $y = \parent(x)$.
If $x$ is the root node, then $\fc(x) = 0$.

Define $\fs(x)$ of a node $x$ as the sum of $\fc(y)$ of all nodes $y$ which come before $x$ in pre-order and are not ancestors of $x$.

By this definition, $N_1 = \fs(z)$ is computed as follows (proof deferred to Section~\ref{sec:proofFS}).
\BL
\label{lem:computeFS}
By maintaining an $\O(n)$-bit structure, we can compute $\fs(x)$ in $\O(1)$ time.
\EL

\item \textbf{Computing $N_2$:}
Note that for any leaf $\ell_j \in \s_2$, $\pLF(j) \leq \pLF(i)$ iff $L[j] \in \Sigma_p$ and either $f_j > f_i$ or $f_j = f_i$ and $j \leq i$.
Therefore, $N_2$ is the number of leaves $\ell_j$ in $\s_2$ which satisfy one of the following conditions: (a) $\pBWT[i] < \pBWT[j] \leq \sigma_p$, or (b) $\pBWT[i] = \pBWT[j]$ and $j \leq i$.
Then, $N_2 = \rangeCount(L_z, R_z,c+1,\sigma_p)+\rangeCount(L_z, i,c,c)$, where $c = \pBWT[i]$.

\medskip

\item \textbf{Computing $N_3$:}
For any leaf $\ell_j \in \s_3$, $\pLF(j) > \pLF(i)$.
Therefore, $N_3 = 0$.

\medskip

\item \textbf{Computing $N_4$:}
Note that $\pBWT[i]$ is same as $(\alphaDepth(v) + 1)$.
Consider a leaf $\ell_j \in \s_4$ with $L[j] \in \Sigma_p$. 
Since the suffix $j$ deviates from the suffix $i$ at the node $v$, we have $f_j \neq f_i$.
Therefore, $\pLF(j) < \pLF(i)$ iff $f_j > f_i$, and the leading character on the path from $v$ to $\ell_j$ is not an s-character.
For a node $x$, $\pc(x)$ is the number of children $y$ of $x$ such that the leading character from $x$ to $y$ is not an s-character.
Note that $\sum_x \pc(x) = \O(n)$. 
Therefore, we encode $\pc(\cdot)$ of all nodes in $\O(n)$ bits using unary encoding, such that $\pc(x)$ can be retrieved in constant time\footnote{
Create a binary string $S$ as follows.
For each node $u$ in a pre-order traversal of $\pST$, append to $S$ a $0$ followed by $\pc(u)$ $1$s.
Append a $0$ at the end, and maintain a $\rank$-$\select$ structure over $S$.
Then $\pc(u)$ for a node $u$, having pre-order rank $k$, is the number of $1$s between the $k$th $0$ and the $(k+1)$th $0$. 
The value is found by using two $\select$ operations followed by two $\rank$ operations. 
}. 
Let $u$ be the $\pc(v)$th child of $v$. 
Then, $N_4$ is the number of leaves $\ell_j$ in $\s_4$ such that $j \leq R_{u}$ and $\sigma_p \geq \pBWT[j] \geq \pBWT[i]$ i.e., $N_4 = \rangeCount( R_z+1, R_u,\pBWT[i],\sigma_p)$.

\end{itemize}
%
%
\begin{algorithm}[!t]
\caption{computes $\pLF(i)$}
\label{alg:LF}
\begin{algorithmic}[1]
\State $c \gets \pBWT[i]$
\If{($c > \sigma_p$)}
 	\State $\pLF(i) \gets 1+\rangeCount(1,n, 1, c-1)+\rangeCount(1,i-1,c,c)$
\Else
	\State $z \gets \zeroNode(\ell_i)$, $v \gets \parent(z)$
 	\State $N_1 \gets \fs(z)$
 	\State $L_z \gets \lmostLeaf(z)$, $R_z \gets \rmostLeaf(z)$
 	\State $N_2 \gets \rangeCount(L_z, R_z,c+1,\sigma_p)+\rangeCount(L_z, i,c,c)$
 	\State $N_3 \gets 0$
 	\If{($\leafc(i)$ is $0$)}
 		\State $u \gets \child(v,\pc(v))$ 
 		\State $N_4 \gets \rangeCount( R_z+1, \rmostLeaf(u),c,\sigma_p)$
 	\EndIf
    \State $\pLF(i) \gets N_1 + N_2 + N_4$
\EndIf
\end{algorithmic}
\end{algorithm}
We summarize the LF mapping procedure in Algorithm~\ref{alg:LF}. 
Once $\alphaDepth(\ell_i)$ is known, $N_1$ is computed in $\O(1)$ time, and both $N_2$ and $N_4$ are computed in $\O(\log \sigma)$ time.
Combining these results with Lemma~\ref{lem:zeronode}, we arrive at Theorem~\ref{thm:lfmapping:a}.

\subsubsection{Proof of Lemma~\ref{lem:computeFS}}
\label{sec:proofFS}
We first present the underlying intuition. 
Since $\sum_x \fc(x)\leq n$, $\fc(\cdot)$ of all nodes can be encoded in $\O(n)$ bits using unary encoding. 
We can decode $\fc(\cdot)$ of any particular node in constant time. 
We carefully select a set of $\O(\frac{n}{\log n})$ nodes, and explicitly storing their $\fs(\cdot)$ values in $\O(n)$ bits. 
The $\fc(\cdot)$ of the remaining nodes can be encoded efficiently using the Method of the Four Russians~\cite{Gusfield1997}. 
Now, we present the details.

We mark nodes in the $\pST$ w.r.t grouping factor $g = \lceil \log n \rceil/\lambda$, where $\lambda$ is a constant to be determined later.
For every prime node $v'$ and a lowest marked node $v^*$, we store $\fs(v')$ and $\fs(v^*)$.
For every marked node $w^*$, we store $\Phi(w^*)$ i.e., the sum of $\fc(w)$ for every $w$ in the subtree rooted at $w^*$.
According to Fact~\ref{fact:marked}, the number of prime nodes and marked nodes is $\O(n/g)$. 
Therefore, the space needed is $\O(\frac{n}{g}\log n) = \O(n)$ bits\footnote{
Maintain a bit-vector $B$ such that $B[k] = 1$ iff the node with pre-order rank $k$ is a prime node w.r.t $g$.
Keep a $\rank$-$\select$ structure over $B$.
In an adjoining array $F$, store the $\fs$ value of each prime node i.e., $F[k]$ stores the $\fs$ value of the prime node corresponding to the $k$th $1$-bit in $B$.
Then, given the pre-order rank of a prime node, we can find its $\fs$ value in $\O(1)$ time using $\rank$-$\select$ operations on $B$ to locate the desired position in $F$.
Likewise, $\fs$ of a lowest marked node and $\Phi$ of a marked node can be retrieved in $\O(1)$ time.
The space needed for each structure is $\O(n)$ bits.}.

Consider a prime node $u'$ and let $u^*$ be its (unique) highest marked descendant.
Let the sequence of nodes on the path from $u'$ to $u^*$ be $u' = u_0,u_1,u_2,\dots,u_k,u_{k+1} = u^*$.
We store the following string corresponding to $u'$:
$$B_{u'} = 1^{n_1} \circ 0 \circ 1^{n_2} \circ 0 \circ \dots \circ 1^{n_k} \circ 0$$
Here, $n_i$ is the number of leaves $j$ in the subtree of $u^*$ such that $$|\nodePath(u_{i-1})| + 2 \leq f_j \leq |\nodePath(u_i)| + 1$$
Maintain a $\rank$-$\select$ structure for $B_{u'}$, and associate it with the prime node $u'$.
For a node $u_i$, $1 \leq i < k$, we define $\Gamma(u_i) = \sum_{i < i' \leq k} n_{i'}$. 
Define $\Gamma(u_k) = 0$.
Observe that $\Gamma(u_i) = \rank_{B_{u'}}(|B_{u'}|,1) - \rank_{B_{u'}}(i',1)$, where $i' = \select_{B_{u'}}(i,0)$, is computed in $\O(1)$ time.
Note that the sum of $|B_{u'}|$ over all prime nodes $u'$ is at most $2n$. 
Therefore, the space needed is $\O(n)$ bits.

\begin{figure}[t]
\centering
\includegraphics[scale=0.8]{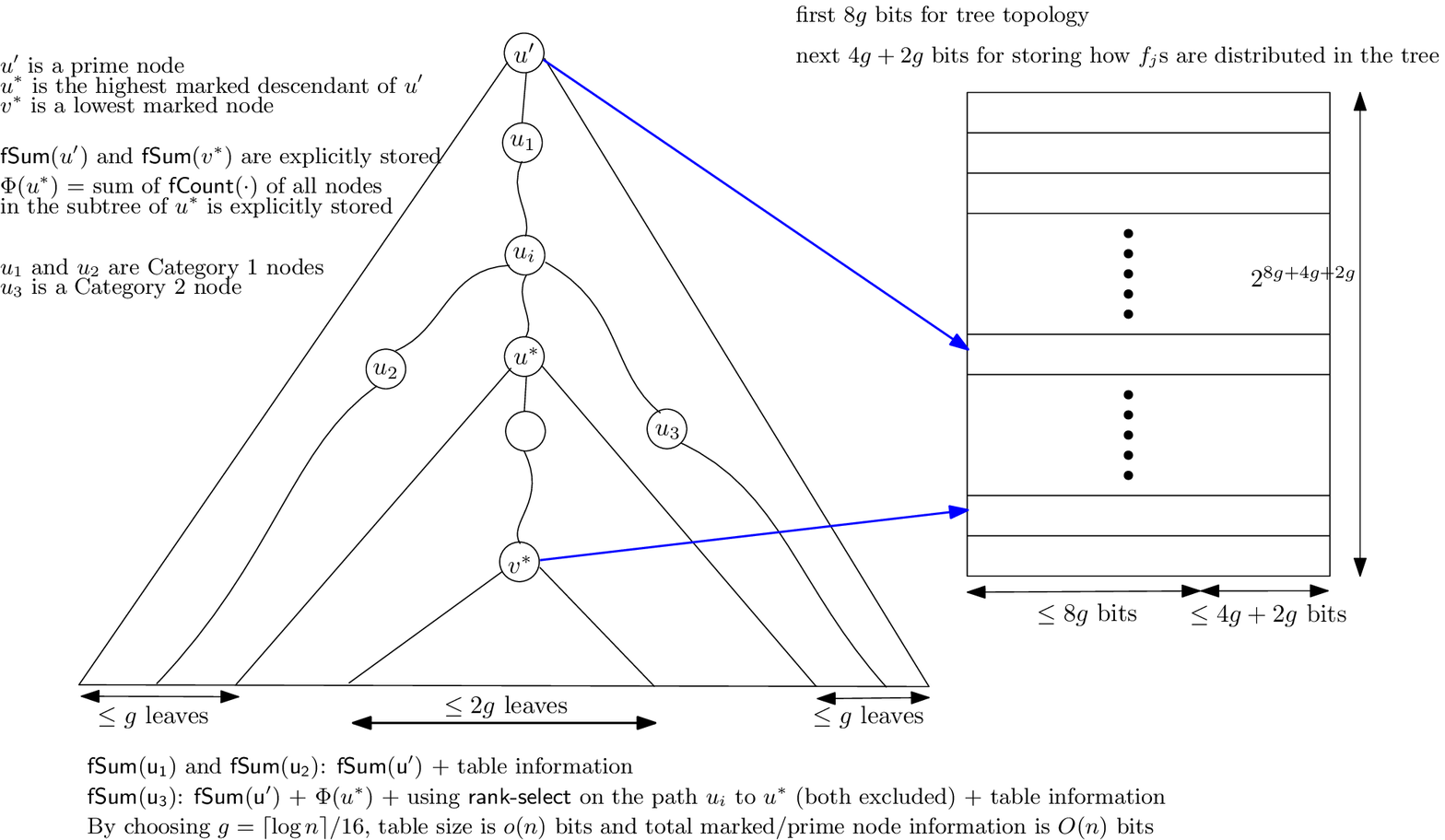}
\caption{Illustration of computation technique of $\fs(\cdot)$}
\label{fig:fSum}
\end{figure}

For any prime node $u'$ and its highest marked descendant $u^*$, the number of nodes in the subtree $\calT$ of $u'$ but not of $u^*$ is at most $4g$  (since $\pST$ is a compacted trie and according to Fact~\ref{fact:marked}).
Therefore, we can encode the subtree $\calT$ in at most $8g$ bits~\cite{SadakaneN10}.
Likewise, the subtree rooted at a lowest marked node can be encoded in at most $8g$ bits.
Now, consider the \emph{local pre-order rank} of any node $v$ in $\calT$. 
Given the pre-order rank of a node in $\pST$, we can easily find its local pre-order rank in constant time.
For every node $v \in \calT$, we classify them into the following two categories (see Figure~\ref{fig:fSum} for an illustration):
\begin{enumerate}
\item[1.] $v$ either lies on or to the left of the $u'$-$u^*$ path i.e., $v$ comes before $u^*$ in pre-order
\item[2.] $v$ lies to the right of the $u'$-$u^*$ path i.e., $v$ comes after $u^*$ in pre-order
\end{enumerate}
For a node $v$ in Category 1, we store $\Psi_1(v) = \fs(v) - \fs(u')$.
For a node $v$ in Category 2, we store $\Psi_2(v) = \fs(v) - \fs(u') - \Phi(u^*) - {\Gamma}(\lca(v,u^*))$.
Consider a node $w$ which lies in $\calT$ but not on the $u'$-$u^*$ path.
The sum of $\fc(w)$ of all such nodes $w$  is at most $2g$ (since the number of leaves in $\calT$ is at most $2g$ according to Fact~\ref{fact:marked}); therefore, maintaining $\Psi_1(v)$ or $\Psi_2(v)$ for each node $v$ requires at most $\lceil \log 2g \rceil$ bits.

We encode $\calT$ by creating a binary string $B_{\calT}$ as follows. 
First $8g$ bits are used to encode the tree topology. 
Now, traverse the tree in pre-order. 
For each node $v$, we append $01^{n_v}$ to $B_{\calT}$, where $n_v$ is the number of leaves $j$ in the subtree of $v$ but not of $u^*$ such that $|\nodePath(\parent(v))| + 2 \leq f_j \leq |\nodePath(v)| + 1$. 
Note that $\sum_{v \in \calT} n_v \leq 2g$ (as there are at most that many leaves). 
Hence, $|B_{\calT}| \leq (8g + 4g+2g) = 14g$.
In an adjoining array, store the value $\Psi_1(v)$ or $\Psi_2(v)$ (depending on which category it belongs to)\footnote{
Consider two trees $\calT_1$ and $\calT_2$ for which $B_{\calT_1} = B_{\calT_2}$.
Note that for Category 1 nodes $v_1$ and $v_2$ (in $\calT_1$ and $\calT_2$ respectively) having the same local pre-order rank, $\Psi_{1,1}(v_1) = \Psi_{1,2}(v_2)$, where $\Psi_{1,i}$ is the  $\Psi_{1}$ function for $\calT_i$.
This is because $\Psi_1(v)$ is only dependent on the string $B_{\calT}$.
Likewise, $\Psi_{2,1}(v_1) = \Psi_{2,2}(v_2)$.}.
The space needed by the array is at most $4g\lceil \log 2g\rceil$ bits.
By maintaining a binary $\rank$-$\select$ on the last $6g$ bits of $B_{\calT}$, we can find the local pre-order rank of $v$, and use it to locate the position of $v$ in the adjoining array. 
The space needed for each $\rank$-$\select$ structure is $o(g)$ bits.
For each prime node, we maintain a pointer to the corresponding tree encoding and the adjoining array.
We choose $\lambda = 16$.
The number of possible strings $B_{\calT}$ is $2^{|B_{\calT}|} = 2^{14 \lceil \log n \rceil/16}$.
Therefore, the space is $2^{14 \lceil \log n \rceil/16}(14g + 4g\lceil \log 2g\rceil + o(g)) = o(n)$ bits.

For a lowest marked node $v^*$ and the tree $\calT'$ rooted at it, we maintain a similar structure.
In this case, apart from the corresponding binary string $B_{\calT'}$ and the adjoining $\rank$-$\select$ structure, for any node $v \in \calT'$, we maintain $\Psi(v) = \fs(v) - \fs(v^*)$, which requires $\lceil \log 2g \rceil$ bits.
Therefore, the space for all such encoding is again $o(n)$ bits.

Summarizing, the total space needed by the data structure is $\O(n)$ bits.
Now, we concentrate on how to find $\fs(v_q)$ for a query node $v_q$.

If $v_q$ is a prime node or a lowest marked node, then we can retrieve $\fs(v_q)$ in $\O(1)$ time.
So, assume otherwise.
We first determine whether it is in the subtree of a lowest marked ancestor, or not. 
(This is checked by determining $v_q$'s lowest marked ancestor and then checking whether that has a marked descendant, or not.)
We have the following two cases.
\begin{itemize}

\item 
If $v_q$ is not in the subtree of a lowest marked node, then we locate the lowest prime ancestor $v_q'$ of $v_q$. 
Now, we locate the unique highest marked node $v_q^*$ corresponding to $v_q'$. 
By comparing the pre-order ranks of $v_q$ and $v_q^*$, we determine whether $v_q$ is Category 1 or 2. 

For Category 1, $\fs(v_q)$ is obtained using $\Psi_1(v_q)$ (stored in the table at $v_q'$) and $\fs(v_q')$ i.e., $\fs(v_q) = \fs(v_q') + \Psi_1(v_q)$.

For Category 2, we have $\fs(v_q) = \Gamma(\lca(v_q,v_q^*)) + \fs(v_q')$ + $\Phi(v_q^*) + \Psi_2(v_q)$.

\item 
Otherwise, we locate the lowest marked ancestor $w_q^*$ of $v_q$. 
Using $\Psi(v_q)$ and $\fs(w_q^*)$, we compute $\fs(v_q)$. 
\end{itemize}
We carry out a constant number of operations in either case, each taking $\O(1)$ time.
This completes the proof of Lemma~\ref{lem:computeFS}.
%

\subsection{A Note on Construction}
Given the p-suffix tree, every component of our index can be constructed in $\O(n\log\sigma)$ time using $\O(n\log n)$ bits of working space. 
Therefore, by first creating $\pST$ using Kosaraju's algorithm~\cite{Kosaraju95}, we have an $\O(n\log\sigma)$ time and $\O(n\log n)$ bit construction algorithm.

\section{Pattern Matching via Backward Search}
\label{sec:backwardSearch}

We use an adaptation of the backward search algorithm in the FM-index~\cite{FerraginaM00}. 
In particular, given a proper suffix $Q$ of $P$, assume that we know the suffix range $[\sp_1,\ep_1]$ of $\prev(Q)$.
Our task is to find the suffix range $[\sp_2,\ep_2]$ of $\prev(c\circ Q)$, where $c$ is the character previous to $Q$ in $P$.

If $c$ is static i.e., $c \in [\sigma_p+1, \sigma]$, then $\prev(c\circ Q)=c\circ \prev(Q)$. 
The backward search in this case is similar to that in FM-index.
Specifically,  
$\sp_2 = 1 + \rangeCount(1,n,1,c-1) + \rangeCount(1,\sp_1-1,c,c)$ and $\ep_2 = \rangeCount(1,n,1,c-1) + \rangeCount(1,\ep_1,c,c)$.
The time required is $\O(\log \sigma)$.

Now, we consider the scenario when $c$ is parameterized.
The idea is to compute the size of the suffix range of $\prev(c\circ Q)$ i.e., $(\ep_2-\sp_2+1)$, and then locate the suffix $\sp_2$.
By maintaining a bit-vector $B[1,\sigma_p]$,  in $\O(|P|)$ time, we first identify all positions $j$, where $P[j] \in \Sigma_p$ is not in $P[j+1,|P|]$. 
We handle the following two cases separately, and summarize the procedure in Algorithm~\ref{alg:backwardSearch}. 

\subsection{Case 1 (c does not appear in $Q$)} 
\label{sec:backwardSearchCase1}

Note that $\pLF(i) \in [\sp_2,\ep_2]$ iff $i \in [\sp_1, \ep_1]$, $L[i]$ is a p-character and $f_i > |Q|$.
This holds iff $i \in [\sp_1, \ep_1]$ and $\pBWT[i] \in [d+1,\sigma_p]$. 
Here, $d$ is the number of distinct p-characters in $Q$, which can be obtained in $\O(1)$ time by initially pre-processing $P$ in $\O(|P|)$ time\footnote{
Maintain a bit-vector $B[1,\sigma_p]$, all values initialized to $0$, and a counter $C$ initialized to $0$.
Scan the text from right to left. 
If a p-character $c_p$ is encountered at position $x$, the required count at $x$ is $C$.
If $B[c_p] = 0$, then increment $C$ by $1$, set $B[c_p]$ to $1$, and continue.}.

Then, $(\ep_2-\sp_2+1) = \rangeCount(\sp_1, \ep_1, d+1, \sigma_p)$ is computed in time $\O(\log \sigma)$. 
Now, $\pLF(i) <\sp_2$ iff $i < \sp_1$, $L[i] \in \Sigma_p$, and $f_i > 1+|\nodePath(\lca(u,\ell_i))|$, where $u = \lca(\ell_{\sp_1}, \ell_{\ep_1})$. 
Therefore, we can compute $\sp_2 = 1+\fs(u)$ in constant time (refer to Lemma~\ref{lem:computeFS}).

\subsection{Case 2 ($c$ appears in $Q$)}
\label{sec:backwardSearchCase2}

Note that $\pLF(i) \in [\sp_2,\ep_2]$  iff $i \in [\sp_1,\ep_1]$, $L[i]$ is a p-character, and $f_i$ is the same as the first occurrence of $c$ in $Q$.
This holds iff $i \in [\sp_1,\ep_1]$ and $\pBWT[i] = d$.
Here, $d$ is the number of distinct p-characters in $Q$ until (and including) the first occurrence of $c$. 
We can compute $d$ in constant time by initially pre-processing $P$ in $\O(|P|\log \sigma)$ time\footnote{
We maintain an array $F[1,\sigma_p]$, all values initialized to $0$, and a balanced binary search tree (initially empty).
Scan the string from right to left, and when a p-character $c_p$ is encountered at a position $x$, check $F[c_p]$. 
If $F[c_p] = 0$, then insert $c_p$ in the tree keyed by $x$.
Otherwise, the count at position $x$ is the number of node in the tree with key at most $F[c_p]$.
Update $F[c_p]$ and the key of $c_p$ in the tree to $x$. 
Since the size of the tree is $\O(\sigma_p)$, search, insertion, and key update time is $\O(\log \sigma)$.}.

Then, $(\ep_2-\sp_2+1) = \rangeCount(\sp_1,\ep_1,d,d)$ is computed in time $\O(\log \sigma)$. 
Consider $i,j \in [\sp_1,\ep_1]$ such that $i<j$ and $\pLF(i),\pLF(j) \in [\sp_2,\ep_2]$.
Now, both $f_i$ and $f_j$ equals the first occurrence of $c$ in $Q$.
Based on Observation~\ref{obs:prevOcc}, we conclude that $\pLF(i) < \pLF(j)$.
Let $i_{min} = \min \{j \mid j \in [\sp_1,\ep_1] \text{ and } \pBWT[j] = d\}$.
Note $i_{min}$ can be computed using the WT over $\pBWT$ in $\O(\log\sigma)$ time.
Then, $\sp_2 =\pLF(i_{min})$ is computed in $\O(\log\sigma)$ time (refer to Theorem~\ref{thm:lfmapping:a}).
%
%
 \begin{algorithm}[!t]
 \caption{computes Suffix Range of $\prev(P[1,p])$}
 \label{alg:backwardSearch}
 \begin{algorithmic}[1]
\State $c \gets P[p]$, $i \gets p$, $P \gets \prev(P)$
\If{($c > \sigma_p$)}
	\State $\sp \gets 1 + \rangeCount(1,n,1,c-1)$
    \State $\ep \gets \rangeCount(1,n,1,c)$
\Else
    \State $\sp \gets 1$
    \State $\ep \gets 1 + \rangeCount(1,n,1,\sigma_p)$
\EndIf
\While{($\sp \leq \ep$ and $i \geq 2$)}
	\State $c \gets P[i-1]$
    \If{($c > \sigma_p$)}
 		\State $\sp \gets 1 + \rangeCount(1,n,1,c-1) + \rangeCount(1,\sp-1,c,c)$
     	\State $\ep \gets \rangeCount(1,n,1,c-1) + \rangeCount(1,\ep,c,c)$ 
     \ElsIf{($c \notin P[i,p]$)}   
     	\State $d \gets$ number of distinct p-characters in $P[i,p]$
     	\State $\sp' \gets \sp$
        \State $\sp \gets 1 + \fs(\lca(\ell_\sp,\ell_\ep))$
     	\State $\ep \gets \sp + \rangeCount(\sp',\ep,d+1,\sigma_p)$
     \Else 
        \State $f \gets$ first occurrence of $c$ in $P[i,p]$
        \State $d \gets$ number of distinct p-characters in $P[i,f]$
     	\State $\sp' \gets \sp$
        \State $\sp \gets \pLF(\min\{j \mid j \in [\sp,\ep] \text{ and } \pBWT[j] = d\})$
     	\State $\ep \gets \sp + \rangeCount(\sp',\ep,d,d)$
     \EndIf
 \EndWhile
 \State \textbf{if} $(\sp < \ep)$ \textbf{then} ``no match found'' \textbf{else return} [$\sp,\ep$]
 \end{algorithmic}
 \end{algorithm}
\\
Summarizing, we find the suffix range of $\prev(P)$ in $\O(|P|\log\sigma)$ time.
Choose $\Delta = \lceil \log_\sigma n \rceil$ in Lemma~\ref{lem:saiSA}. 
According to Theorem~\ref{thm:lfmapping:a}, we have $t_{\pLF} = O(\log \sigma)$.
Theorem~\ref{thm:param:a} is now immediate.


\section{Squeezing to Succinct Space}
\label{sec:succinctSpace}

In this section, we prove Theorem~\ref{thm:lfmapping:b}, and then use it to prove Theorem~\ref{thm:param:b}.
To achieve the space bound of Theorem~\ref{thm:lfmapping:b}, we first need an alternative way (to that of Lemma~\ref{lem:zeronode}) for computing $\zeroNode$. 
To this end, we present the following lemma.
\BL 
\label{lem:zeronode2}
We can find $\zeroNode(\ell_i)$ in $\O(\log \sigma \log \log \sigma)$ time using the WT over $\pBWT$ and an additional $\O(n)$-bit structure.
\EL
Based on Section~\ref{sec:lfMapping} and the above lemma, Theorem~\ref{thm:lfmapping:b} is immediate.
Therefore, we can find the suffix range of $\prev(P)$ in $\O(|P|\log \sigma \log \log \sigma)$ time (see Section~\ref{sec:backwardSearch}).
The additional $\O(n\log \sigma)$ space (above the $n\log \sigma$ space needed for the WT over $\pBWT$) in Theorem~\ref{thm:param:a} is due to the choice of  $\Delta = \lceil \log_\sigma n\rceil$ in Lemma~\ref{lem:saiSA}.
We can improve this space to $\O(n)$ bits by choosing $\Delta = \lceil \log n \rceil$. 
Reporting a suffix array value requires $\O(\log n\log \sigma \log \log \sigma)$ time.
Theorem~\ref{thm:param:b} follows immediately.

The rest of this section is dedicated for proving Lemma~\ref{lem:zeronode2}.
Before, getting into the details, we first present the key intuition.
Let $u_1$ and $u_2$ be two consecutive nodes on the path from root to $\ell_i$, and $\alpha$ be the number of leaves $\ell_j$ in the subtree of $u_2$ such that $f_j$ lies on the edge $(u_1,u_2)$.
Then, if we know $\alphaDepth(u_1)$ or the largest $\pBWT[j] \leq \alphaDepth(u_1)$ for a leaf $\ell_j$ in the subtree of $u_1$, we can find $\beta$ i.e., the largest $\pBWT[j'] \leq \alphaDepth(u_2)$ value for a leaf $\ell_{j'}$ under the subtree of $u_2$ using a $\rangeNV$-query.
By comparing $\beta$ with $\pBWT[i]$, we know whether $u_2 = \zeroNode(\ell_i)$, or not.
Note that the $\alpha$ values of all nodes in the tree can be encoded in $\O(n)$ bits using unary encoding, such that each $\alpha$ value can be decoded in $\O(1)$ time.
By storing the $\alphaDepth$ of a carefully selected set of $2\lceil \frac{n}{\log \sigma}\rceil$ marked nodes, we can find a marked node $w$, which is at most $\log \sigma$ nodes above $\zeroNode(\ell_i)$.
Use $\rangeNV$ queries at each node on the path from $w$ to $\ell_i$ to find $\zeroNode(\ell_i)$.
The space-and-time tradeoff is $\O(n)$ bits vs $O(\log^2 \sigma)$ time; details are presented in Section~\ref{sec:succinctPrelim}.
A more sophisticated scheme in Section~\ref{sec:succinctFinal} enables us to achieve the claimed bounds.
We begin with the following lemma.
\BL
\label{lem:lowestMarked}
By maintaining the WT over $\pBWT$ and an additional $\O(n)$-bit structure, in $\O(\log \sigma)$ time, we can find the lowest marked ancestor $w_i$ of a leaf $\ell_i$ such that $\alphaDepth(w_i) < \pBWT[i]$.
\EL
\begin{proof}
%
%
Identify marked nodes with  $g=\lceil \log\sigma \rceil$.
Maintain a bit-array $B$ such that $B[k] = 1$ iff the node with pre-order rank $k$ is a marked node. 
Also, maintain a $\rank$-$\select$ structure on $B$.
The space needed is $\O(n)$ bits. 
We also maintain an array $D$, such that $D[k]$ equals the $\alphaDepth$ of the marked node corresponding to the $k$th $1$-bit in $B$. 
Given a marked node with pre-order rank $k'$, its corresponding position in $D$ is given by $\rank_B(k',1)$.
We do not maintain $D$ explicitly; instead, we maintain a wavelet tree over it.
The space needed is $\O(\frac{n}{g} \log \sigma) = \O(n)$ bits.

Given a leaf node $\ell_i$, in constant time, first locate the lowest marked ancestor $u$ (refer to Fact~\ref{fact:marked}).
Then, find the position $j$ corresponding to $u$ in the array $D$.
If $\alphaDepth(u) < \pBWT[i]$, then $w_i = u$, and we are done.
Otherwise, locate $j' = \pred_D(j,\pBWT[i])$ i.e., the rightmost position $j' < j$ in $D$ such that $D[j'] < \pBWT[i]$; this is achieved in $\O(\log \sigma)$ time.
(Since the root node is marked, and its $\alphaDepth$ equals $0$, the position $j'$ necessarily exists.)
Obtain the marked node $v$ corresponding to the $j'$th $1$-bit in $B$ via a $\select_B(j',1)$ operation.
Then, $w_i = \lca(u,v)$.
The time required is $\O(\log \sigma)$.

To see the correctness first observe that $\lca(u,v)$ is marked, and is an ancestor of $\ell_i$. 
Also, for any node $x$, $\alphaDepth(x) \geq \alphaDepth(\parent(x))$.
Thus, $\alphaDepth(\lca(u,v)) \leq \alphaDepth(v) < \pBWT[i]$.
If $\lca(u,v)$ is not the desired node, then it has a marked descendant $u' \neq u$ on the path to $u$ that satisfies $\alphaDepth(u') < \pBWT[i]$. 
But then $u'$ appears after $v$ and before $u$ in pre-order, a contradiction.
\qed
\end{proof}

\subsection{An $\O(n)$-bit and $\O(\log^2 \sigma)$-time Index}
\label{sec:succinctPrelim}

\begin{figure}[t]
\centering
\includegraphics[scale=0.8]{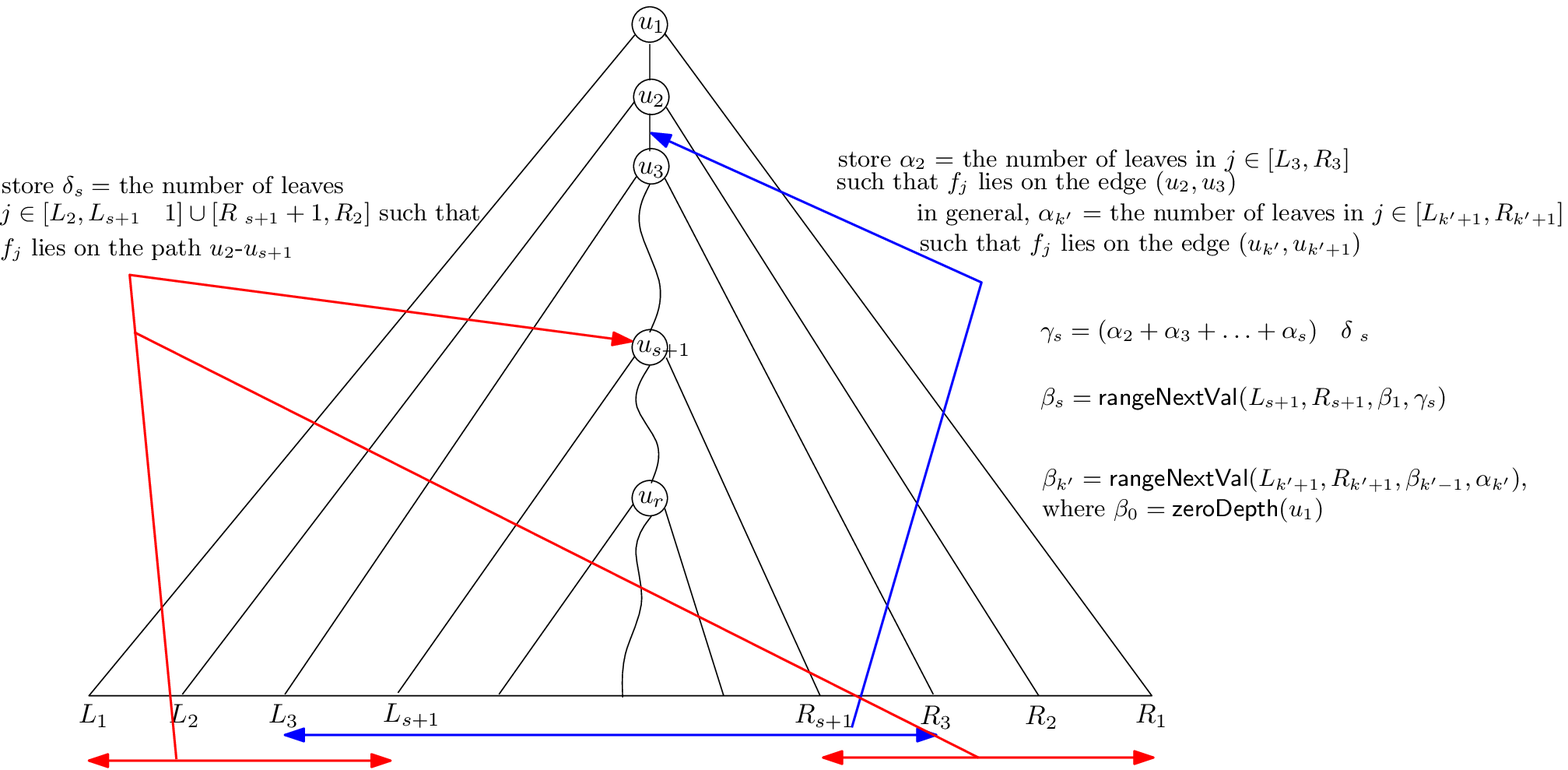}
\caption{Illustration of finding $\zeroNode$ using $\O(n)$ bits}
\label{fig:zeroNode2}
\end{figure}

Mark nodes w.r.t a parameter $g = \lceil \log \sigma \rceil$. 
To verify whether a node is marked (w.r.t $g$), we maintain a bit-vector $B$ where an entry $j$ is set to $1$ iff the node with pre-order rank $j$ is marked.
(We may simply use the bit-vector $B$ of Lemma~\ref{lem:lowestMarked}.)
Using Lemma~\ref{lem:lowestMarked}, we first identify the lowest marked ancestor $w_i$ of $\ell_i$ that satisfies $\alphaDepth(w_i) < \pBWT[i]$ in $\O(\log \sigma)$ time.
Consider the nodes $w_i = u_1, u_2, \dots, u_k = \ell_i$ on the $w_i$-$\ell_i$ path.
For any node $u_{k'}$, $1 \leq k' \leq k$, denote by $[L_{k'},R_{k'}]$ the range of leaves in its subtree.
See Figure~\ref{fig:zeroNode2} for an illustration.
Let $\alpha_{k'}$, $k' < k$,  be the number of leaves $j \in [L_{k'+1},R_{k'+1}]$ that satisfy $\alphaDepth(u_{k'}) < \pBWT[j] \leq \alphaDepth(u_{k'+1})$.
Using the WT over $\pBWT$, we find $\beta_1$ = the $\alpha_1$th smallest value in $\pBWT[L_2,R_2]$ that is greater than $\alphaDepth(u_1)$ i.e., $\beta_1 = \rangeNV(L_2,R_2,\alphaDepth(u_1),\alpha_1)$.
If $\pBWT[i] \leq \beta_1$, then clearly $u_2 = \zeroNode(\ell_i)$.
Otherwise, by the definition of $\alpha_1$ and $\beta_1$, we must have $\pBWT[i] > \alphaDepth(u_2) \geq \beta_1$.
We again query to obtain $\beta_2 = \rangeNV(L_3,R_3,\beta_1,\alpha_2)$.
Observe that crucially $\beta_2$ is the same as $\rangeNV(L_3,R_3,\alphaDepth(u_2),\alpha_2)$.
We again compare $\pBWT[i]$ and $\beta_2$ to determine whether $u_3= \zeroNode(\ell_i)$, or not; in the latter case, we query again.
Repeat this process until we find $\zeroNode(\ell_i)$.
By the definition of $w_i = u_1$, any marked node $u_r$ on the path from $u_2$ to $\ell_i$ must satisfy $\alphaDepth(u_r) \geq \pBWT[i]$, and we will find $\zeroNode(\ell_i)$ by the time we  reach $u_r$.
(Note that $\zeroNode(\ell_i)$ may be the same as $u_r$.)
If $u_1$ is a lowest marked node, then we will find $\zeroNode(\ell_i)$ by the time we reach $\ell_i$. 
(Note that $\zeroNode(\ell_i)$ may be the same as $\ell_i$.)
In either case, the number of nodes traversed is at most $g$ (refer to Fact~\ref{fact:marked}).
Therefore, the time required is $\O(g\log \sigma) = \O(\log ^2 \sigma)$.

In order to find $\alpha_{k'}$, first observe that each leaf $\ell_j$ contributes to exactly one $f_j$ on a certain edge, which in turn contributes exactly one to the $\alpha_{k'}$ value for this edge.
Therefore, crucially, the total of all $\alpha_{k'}$ values over the entire tree is at most $n$. 
Using unary encoding, the values over all edges can be stored in $\O(n)$ bits, and the value of an edge can be accessed in constant time
\footnote{
Create a binary string $S$, initialized to $0$, as follows.
For each non-root node $u$ in a pre-order traversal of $\pST$, append to $S$ as many $1$s as the value of $\alpha_{k'}$ on the edge to its parent, followed by a $0$. 
Maintain a $\rank$-$\select$ structure over $S$.
The value of an edge $(\parent(u),u)$, where $u$ has pre-order rank $k > 1$, is the number of $1$s between the $k$th $0$ and the $(k-1)$th $0$. 
The value is found by using two $\select$ operations followed by two $\rank$ operations.}.
This completes the proof of the $\O(n)$ space and $\O(\log^2 \sigma)$ time solution.

\subsection{Improving the time to $\O(\log \sigma \log \log \sigma)$}
\label{sec:succinctFinal}

To improve the query time, we maintain another group of marked nodes w.r.t $g' = \lceil \log \log \sigma \rceil$. 
Also maintain a bit-vector $B'$ to identify whether a node is marked, or not.
Now, given the lowest $g'$-marked ancestor $w_i'$ of $\ell_i$ that satisfies $\alphaDepth(w_i') < \pBWT[i]$, we now need to traverse and query at  $\leq g'$ nodes. 
Thus, the time improves to $\O(\log \sigma \log \log \sigma)$.
However, we need an alternative way to find the node $w_i'$ as maintaining their $\alphaDepth$ in a WT (as in Lemma~\ref{lem:lowestMarked}) will need $\O(\frac{n}{g'}\log \sigma) = \O(n\log \sigma / \log \log \sigma)$ bits, which violates the space bound.

We can find $\beta_1$ (corresponding to $u_2$ as described above) with a single query costing $\O(\log \sigma)$ time.
For any $k'$, $1 < k' < k$, the successive queries starting from $u_3$ to $u_{k'+1}$ gives us $\beta_{k'}$. 
Observe that $\beta_{k'}$ is exactly the $\gamma_{k'}$th smallest number in $\pBWT[L_{k'+1},R_{k'+1}]$ which is greater than $\beta_1$, where $\gamma_{k'}$ equals the number of leaves $j \in [L_{k'+1},R_{k'+1}]$ that satisfy $\beta_1 < \pBWT[j] \leq \alphaDepth(u_{k'+1})$.
Alternatively, $\gamma_{k'}$ equals the number of leaves $j \in [L_{k'+1},R_{k'+1}]$ such that $f_j$ lies on the path from $u_2$ to $u_{k'+1}$.
Therefore, we can find $\beta_{k'}$ with a single $\rangeNV(L_{k'+1},R_{k'+1},\beta_1,\gamma_{k'})$-query, provided that the value of $\gamma_{k'}$ is known.

Consider the traversal from $u_2$ to $u_k = \ell_i$.
If $u_1$ is a lowest marked node, then $|\Leaf(u_2)\setminus \Leaf(u_{k'})| \leq 2g$, $1 < k' \leq k$ (refer to Fact~\ref{fact:marked}).
Otherwise, $u_2$ is a prime node; as before, let $u_r$ be the highest $g$-marked node on the path below $u_1$.
We have $\alphaDepth(u_r) \geq \pBWT[i]$, and $|\Leaf(u_2)\setminus \Leaf(u_{k'})| \leq 2g$, $1 < k' < r$ (refer to Fact~\ref{fact:marked}).
If $r \leq g'$, then we can directly use the solution of Section~\ref{sec:succinctPrelim}; query time will be $\O(\log \sigma \log \log \sigma)$ as we will carry out at most $g'$ $\rangeNV$-queries. 
If $r > g'$, then we will encounter at least one $g'$-marked node in the traversal (refer to Fact~\ref{fact:marked}).
Among these $g'$-marked nodes lies the node $w'_i$, which we intend to find.

We show the scenario when $u_r$ exists; the case where $u_1$ is a lowest marked node follows similarly.
Let us first concentrate on how to find $\gamma_s$ for a $g'$-marked node $u_{s+1}$ on the path from $u_2$ to $u_{r-1}$.
Consider the leaves in $\mathcal{L}_s = \Leaf(u_2)\setminus \Leaf(u_{s+1})$. 
Among these at most $2g$ leaves, we maintain $\delta_s = $ the number of leaves $j \in \mathcal{L}_s$ for which $f_j$ lies on the path from $u_2$ to $u_{s+1}$.
Note that $\delta_s \leq 2g = 2\lceil \log \sigma \rceil$ can be maintained in $\O(\log \log \sigma)$ bits.
Thus, the space for maintaining $\delta_s$ values over all $g'$-marked nodes is bounded by $\O(\frac{n}{g'}{\log \log \sigma}) = \O(n)$ bits.
(Basically, for each $g'$-marked node we store the corresponding $\delta$ value w.r.t its lowest $g$- prime ancestor.)
By definition, $\gamma_s = (\alpha_2 + \alpha_3 + \dots + \alpha_s) - \delta_s$.
The summation part of $\gamma_s$ is obtained by walking on the $u_2$-$u_{s+1}$ path and querying at each edge (which requires $\O(1)$ time per edge using the unary encoding as discussed before).
Since at most $g$ nodes are encountered, the time needed is $\O(\log \sigma)$.

We are now equipped to find the desired node $w'_i$.
Traverse the path from $u_2$ to $\ell_i$ (using $\la(\ell_i,\cdot)$-queries) and keep track of the $g'$-marked nodes encountered until we hit the $g$-marked node $u_r$ (using the bit-vectors $B$ and $B'$).
Now, we can binary search on this $g'$-marked nodes to find $w'_i$.
Specifically, for any $g'$-marked $u_{s+1}$, we first find $\gamma_s$ and then use it to find $\beta_s$; the time required is $\O(\log \sigma)$.
Now, by comparing $\beta_s$ with $\pBWT[i]$, we know whether we should proceed to the right (i.e., below $u_{s+1}$) or to the left (i.e., above $u_{s+1}$) to locate $w'_i$.
The number of searches is $\O(\log g) = \O(\log \log \sigma)$. 
Therefore, the total time needed is $\O(\log \sigma \log \log \sigma)$, as desired.

\section{Structural Suffix Trees}
\label{sec:struct}

Shibuya considered the following variant of Problem~\ref{problem}, known as \emph{Structural Matching.}

\BDE[\cite{Shibuya00}]
\label{prob:struct}
Let $\Sigma$ be an alphabet of size $\sigma \geq 2$, which is the union of two disjoint sets: $\Sigma_s$ having $\sigma_s$ static characters (s-characters) and $\Sigma_p$ having $\sigma_p$ parameterized characters (p-characters).  
For each p-character, we associate a p-character (called the complement character).
Two strings $S$ and $S'$, having the same length, are a structural-match (s-match) iff 
\begin{itemize}
\item $S$ and $S'$ are a p-match, and
\item if a p-character $x$ in $S$ is renamed to $y$ in $S'$, then the complement (if exists) of $x$ in $S$ is renamed to the complement of $y$ in $S'$.
\end{itemize}
Let $\T$ be a text having $n$ characters chosen from $\Sigma$. 
We assume that $\T$ terminates in an s-character $\$$ that appears only once.
The task is to index $\T$, such that for a pattern $P$ (also over $\Sigma$), we can report all  starting positions (occurrences) of the substrings of $\T$ that are an s-match with $P$.
\EDE
Consider the following examples.
Let $\Sigma_s = \{A,B,C\}$ and $\Sigma_p = \{w,x,y,z\}$, where the complement pairs are $x$-$w$ and $y$-$z$.
Then $AxByCx$ is an s-match with $AyBxCy$; in this case, there are no complementarity requirements i.e., it is simply a p-match.
Also, $AxBwCx$ is an s-match with $AzByCz$; here, $x$ is paired with $z$, and $w$ (complement of $x$) is paired with $y$ (complement of $z$).
However, $AxBwCx$ is not an s-match with $AzBxCz$ (even though they are a p-match); this is because as $x$ is paired with $z$, $w$ should have been paired with $y$.

Shibuya presented a $\Theta(n\log n)$-bit and $\O(|P|\log \sigma + occ)$ time index for this problem.
We present the following new results.
%
%
\BT
\label{thm:structural}
All $occ$ positions where $P$ is an s-match with $\T$ can be found as follows: \emph{(a)} in $\O(|P|\log \sigma + occ\cdot \log n)$ time using an $\O(n\log \sigma)$-bit index, and \emph{(b)} in $\O((|P| + occ\cdot \log n) \log \sigma \log \log \sigma)$ time using an $n\log \sigma + \O(n)$-bit index.
\ET
%
%
We encode a string $S$ as $\compl(S)$ as follows.
If $S[i]$ is static, then $\compl(S)[i] = S[i]$.
Consider a p-character $S[i]$ and let $j^+ < i$ and $j^- < i$ be the rightmost occurrence of $S[i]$ and the complement of $S[i]$ in $S[1,i-1]$.
If there is no occurrence $j^+$ (resp. $j^-$), we let $j^+ = -1$ (resp. $j^- = -1$).
If $j^+ = j^- =-1$, then replace $S[i]$ by $0$.
Otherwise, if $j^+ > j^-$, then $\compl(S)[i] = (i-j^+)$.
Otherwise, if $j^- > j^+$, then $\compl(S)[i] = -(i-j^-)$.
For example, $\compl(AxByCx) = A0B0C4$ and $\compl(AxBwAwCxAx) = A0B(-2)A2C(-2)A2$.
The following lemma is trivially true.
\BL
\label{lem:compl}
Two strings $S$ and $S'$ are an s-match iff $\compl(S) = \compl(S')$.
\EL
The structural suffix tree ($\tST$) is the compact trie of all the strings in the set $\{\compl(\T[i,n] \mid 1 \leq i \leq n\}$.
We analogously define the \emph{locus} and the \emph{suffix range} of a pattern in this case.

Consider each circular suffix of $\T$. 
Sort them based on their $\compl(\cdot)$ encoding, where character precedence is determined by Convention~\ref{convention:lexico}.
Then, obtain the last character $L[i]$ of the $i$th lexicographically smallest suffix.
Define $\tLF(i) = \tSAI[\tSA[i]-1]$, where $\tSA[\cdot]$ and $\tSAI[\cdot]$ denotes the suffix array and inverse suffix array value according to $\compl(\cdot)$.
Denote by $f_i^+$ (resp. $f_i^-$) the first occurrence of $L[i]$ (resp. the complement of $L[i]$) in the circular suffix $\T_i$.
In case, there is no occurrence of $L[i]$'s complement, we take $f_i^- = n+1$.
We define the structural BWT as follows.
\begin{equation*}
\tBWT[i] =
\begin{cases}
L[i], & \text{ if } L[i] \in \Sigma_s,
\\
\text{number of distinct p-characters in }\T_{\tSA[i]}[1,f_i^+],& \text{ if }  L[i] \in \Sigma_p \text{ and }f_i^+ < f_i^-,
\\
- \text{number of distinct p-characters in }\T_{\tSA[i]}[1,f_i^-],& \text{ if }  L[i] \in \Sigma_p \text{ and }f_i^+ > f_i^-.
\end{cases}
\end{equation*}
\BO 
\label{obs:complOcc} 
For any $1\leq i \leq n$, $\compl(\T_{\pSA[i]-1})=$
\begin{equation*}
\begin{cases}
\tBWT[i] \circ \compl(\T_{\tSA[i]})[1,n-1],
 & \text{if } \tBWT[i] \in \Sigma_s,\\
0 \circ \compl(\T_{\tSA[i]})[1,f_i^+-1] \circ f_i^+ \circ \compl(\T_{\tSA[i]})[f_i^++1,n-1], &   \text{if } \tBWT[i] \in [1,\sigma_p]\\
0 \circ \compl(\T_{\tSA[i]})[1,f_i^- -1] \circ -f_i^- \circ \compl(\T_{\tSA[i]})[f_i^- +1,n-1], &   \text{if } \tBWT[i] \in [-\sigma_p, -1]
\end{cases}
\end{equation*}
\EO
It follows from Observation~\ref{obs:complOcc} that for two suffixes $i$ and $j$, of which at least one is preceded by an s-character, determining whether $\tLF(i) < \tLF(j)$, or not, remains same as in Lemma~\ref{lem:atleastOneStatic}.
So, we concentrate on the scenario where both $L[i]$ and $L[j]$ are p-characters.

\BL 
\label{lem:bothParamStruct}
Assume $i<j$ and both $L[i]$ and $L[j]$ are parameterized. 
Let $u$ be the lowest common ancestor of $\ell_i$ and $\ell_j$ in $\pST$, and $z$ be the number of $0$'s in the string $\nodePath(u)$.
Then,
\begin{enumerate}[label={\emph{(\arabic*)}}]

\item
If $|\tBWT[i]|, |\tBWT[j]| \leq z$, then $\tLF(i) < \tLF(j)$ iff one of the following holds:
	\begin{enumerate}[label={\emph{(\alph*)}}]
	\item $\tBWT[i], \tBWT[j] > 0$ and $\tBWT[i] \geq \tBWT[j]$
	
	\item $\tBWT[i] < 0 < \tBWT[j]$
	
	\item $\tBWT[i], \tBWT[j] < 0$ and $|\tBWT[i]| \leq |\tBWT[j]|$
	
	\end{enumerate}
\item 
If $|\tBWT[i]| \leq z < |\tBWT[j]|$, then $\tLF(i) < \tLF(j)$ iff $\tBWT[i] < 0$

\item
If $|\tBWT[i]| > z \geq |\tBWT[j]|$, then $\tLF(i) < \tLF(j)$ iff $\tBWT[j] > 0$

\item If $|\tBWT[i]|, |\tBWT[j]| > z$, then $\tLF(i) > \tLF(j)$ iff one of the following holds:
	\begin{enumerate}[label={\emph{(\alph*)}}]
	\item $\tBWT[i] = z+1$, the leading character on the path from $u$ to $\ell_i$ is $0$, and the leading character on the path from $u$ to $\ell_j$ is not an s-character.
	
	\item $\tBWT[j] = -(z+1)$, and the leading character on the path from $u$ to $\ell_j$ is $0$.
	
	\end{enumerate}

\end{enumerate}
\EL
The $\tBWT$ can be maintained in $n\log \sigma + \O(n)$ bits to support the same operations outlined in Section~\ref{sec:WT} in $\O(\log \sigma)$ time\footnote{ 
Create an array $\tBWT'$ such that $\tBWT'[i] = \tBWT[i] + \sigma_p$.
The alphabet size is at most $\sigma + \sigma_p \leq 2\sigma$.
Therefore, the WT over $\tBWT'$ occupies space $n\log (2\sigma) + o(n) = n \log \sigma + \O(n)$ bits.
A query on $\tBWT$ can be easily answered by the same query on $\tBWT'$ with $\sigma_p$ added to the relevant parameters.
Time needed is $\O(\log (2\sigma))  = \O(\log \sigma)$.}.

\subsection{Redefining $\zeroNode(\ell_i)$}
For a leaf $\ell_i$, $\tBWT[i] \in [1,\sigma_p]$, we define $\zeroNodePos(\ell_i)$ to be the locus of $\nodePath(\ell_i)[1,f_i^+]$. 
Equivalently, $\zeroNodePos(\ell_i)$ is the highest node $w$ on the root to $\ell_i$ path such that $\alphaDepth(w) \geq \tBWT[i]$.
Likewise, for a leaf $\ell_i$, $\tBWT[i] < 0$, $\zeroNodeNeg(\ell_i)$ is the locus of $\nodePath(\ell_i)[1,f_i^-]$.
Equivalently, $\zeroNodeNeg(\ell_i)$ is the highest node $w$ on the root to $\ell_i$ path such that $\alphaDepth(w) \geq -\tBWT[i]$.

To locate $\zeroNodePos(\ell_i)$ or $\zeroNodeNeg(\ell_i)$, it suffices to find the lowest node $u$ on the root to $\ell_i$ path such that $\alphaDepth(u) < |\tBWT[i]|$.
(The desired node $w$ is the child of $u$ on this path.)

First, let us concentrate on modifying Lemma~\ref{lem:zeronode}.
We locate the node $v$ with pre-order rank $\pred_D(j,|\tBWT[i]|)$ query;  here, $D[k]$ equals the $\alphaDepth$ of the node with pre-order rank $k$, and $j$ is the pre-order rank of $\ell_i$. 
Then $u = \lca(\ell_i,v)$. 
The correctness and complexity claims remain the same as in the case of Lemma~\ref{lem:zeronode}.

Things are a little bit trickier when it comes to modifying Lemma~\ref{lem:zeronode2}.
First of all, we have to find a $g$-marked node $w_i$ of $\ell_i$, where $g = \lceil \log \sigma \rceil$, such that $\alphaDepth(w_i) < |\tBWT[i]|$.
This part is achieved by simply replacing the query $\pred_D(j,\pBWT[i])$ in the proof of Lemma~\ref{lem:lowestMarked} with the query $\pred_D(j,|\tBWT[i]|)$.
The correctness and complexity claims remain the same as in the case of Lemma~\ref{lem:lowestMarked}.
We omit details and definition of the notations in what follows, as they are either the same or follow immediately from the proof of Lemma~\ref{lem:zeronode2}.

As before, $\beta_0 = \alphaDepth(w_i)$. 
For $k' > 0$, $\beta_{k'}$  is the $\alpha_{k'}$th smallest $|\tBWT[j]|$ value, which is greater than $\beta_{k'-1}$, $L_{k'+1} \leq j \leq R_{k'+1}$.
Here, $\alpha_{k'}$ is the number of leaves $j$, $L_{k'+1} \leq j \leq R_{k'+1}$, such that either $f_j^+$ or $f_j^-$ (depending on whether $\pBWT[j] \in [1,\sigma_p]$ or $\pBWT[j] < 0$) lies on the edge from $u_{k'+1}$ to its parent.
In the current scenario, starting from $u_2$, we have to find the first $\beta_{k'} \geq |\tBWT[i]|$ corresponding to a node $u_{k'+1}$ on the path $w_i = u_1, u_2, \dots, u_k = \ell_i$.

We write $\alpha_{k'}$ separately as $\alpha_{k'}^+$ and $\alpha_{k'}^-$ (corresponding to $f_j^+$ and $f_j^-$).
All these values can be maintained using unary encoding in $\O(n)$ bits, such that a $\alpha_{k'}^+$ or $\alpha_{k'}^-$ value can be retreived in $\O(1)$ time.
Observe that crucially the desired value $\beta_{k'}$ is the maximum of the  $\alpha_{k'}^+$th smallest value greater than $\beta_{k'-1}$ or the absolute value of the $\alpha_{k'}^-$th largest value smaller than $-\beta_{k'-1}$.
We find $\beta_{k'}^+ = \rangeNV(L_{k'+1}, R_{k'+1}, \beta_{k'-1}, \alpha_{k'}^+)$ and $\beta_{k'}^- = \rangePV(L_{k'+1}, R_{k'+1}, -\beta_{k'-1}, \alpha_{k'}^-)$.
Then, $\beta_{k'} = \max\{\beta_{k'}^+,-\beta_{k'}^-\}$ is obtained in time $\O(\log \sigma)$.

Based on the above discussion we conclude, $\beta_{k'}^+ = \rangeNV(L_{k'+1}, R_{k'+1}, \beta_1, \gamma_{k'}^+)$ and $\beta_{k'}^- = \rangePV(L_{k'+1}, R_{k'+1}, -\beta_1, \gamma_{k'}^-)$, where $\gamma_{k'}^+$ (resp. $\gamma_{k'}^-$) is the number of leaves $j \in [L_{k'+1}, R_{k'+1}]$ such that $f_j^+$ (resp. $f_j^-$) lies on the path from $u_2$ to $u_{k'+1}$.
Now, define $\delta^+_{k'}$ as the number of leaves $j$ in $\mathcal{L}_{k'} = \Leaf(u_2) \setminus \Leaf(u_{k'+1})$ such that $f_j^+$ lies on the path from $u_2$ to $u_{k'+1}$.
Likewise, define $\delta^-_{k'}$.
Both can be maintained in $\O(\log \log \sigma)$ bits at each $g'$-marked node, where $g' = \lceil \log \log \sigma \rceil$. 
The total space needed is $\O(n)$ bits.
Now, for a $g'$-marked node on the path from $w_i$ to $u_{k'+1}$, $\gamma_{s}^+ = (\alpha_2^+ + \alpha_3^+ + \cdots + \alpha_{s}^+) - \delta^+_{s}$ is obtained by walking on the path and querying at each node.
Likewise, we can find $\gamma_{s}^-$.
The time needed is $\O(\log \sigma)$ as we need to walk at most $\lceil \log \sigma \rceil$ nodes.
Therefore, as in Lemma~\ref{lem:zeronode2}, we can find the desired $g'$-marked node via a binary search in time $\O(\log \sigma \log \log \sigma)$. 

Summarizing, we have the following lemma.
\BL
\label{lem:zeroStruct}
For any leaf $\ell_i$, where $\tBWT[i] \in [1,\sigma_p]$, we can compute $\zeroNodePos(\ell_i)$ as follows: \emph{(a)} in $\O(\log \sigma)$ time using an $\O(n \log \sigma)$-bit structure, and \emph{(b)} in $\O(\log \sigma \log \log \sigma)$ time using an $n \log \sigma + \O(n)$-bit structure.
Likewise, for any leaf $\ell_i$, where $\tBWT[i] < 0$, we can compute $\zeroNodeNeg(\ell_i)$ is the same space-and-time tradeoffs.
\EL

\subsection{Redefining $\fs(x)$}

Define $\fcPos(x)$ as the number of leaves $\ell_j$ under the subtree of $x$, $\tBWT[j] \in [1,\sigma_p]$, such that $|\nodePath(y)| + 2 \leq f_j \leq |\nodePath(x)| + 1$, where $y = \parent(x)$.
Define $\fsPos(x)$ as the sum of $\fcPos(y)$ of all nodes $y$ which come before $x$ in pre-order and are not ancestors of $x$.


Now, we define $\fsNegRev(x)$ as the number of leaves $\ell_j$, where $\tBWT[j] < 0$, that appear after $\rmostLeaf(x)$ in pre-order and satisfy $f_j^- > 1 + |\nodePath(\lca(x,\ell_j))|$.
Thus, $\fsNegRev(x)$ is the sum of $\fcNeg(y)$ of all nodes $y$ which come after $\rmostLeaf(x)$ in pre-order (and by definition, are not ancestors of $x$).

Recall that in the original computation of $\fs(x)$, we only encode the information for those leaves $\ell_j$, where $\pBWT[j] \in [1,\sigma_p]$, i.e., we simply ignore all other leaves.
Therefore, while creating the $\fs$-computation structure, by only considering the leaves $\ell_j$, where $\tBWT[j] \in [1,\sigma_p]$, we can obtain a structure for computing  $\fsPos(x)$.
Intuitively, $\fsNegRev(x)$ is same as $\fsPos(x)$ with the following modifications:
\begin{itemize}
\item We consider the leaves $\ell_j$ for which $\tBWT[j] < 0$.
\item The leaves in $\tST$ are ordered in reverse lexicographic order to that of  Convention~\ref{convention:lexico}.
\end{itemize}
Therefore, we can compute $\fsNegRev(x)$ using a structure similar to that for computing $\fsPos(x)$. 

Summarizing, we have the following lemma.
\BL
\label{lem:fSumStruct}
For any node $x$, we can compute $\fsPos(x)$ and $\fsNegRev(x)$ in $\O(1)$ time by maintaining an $\O(n)$-bit structure.
\EL

\subsection{Implementing $\tLF$ Mapping}
\label{sec:tLFMapping}

If $\tBWT[i] > \sigma_p$, then $$\LF(i) = 1 + \rangeCount(1,n,-\sigma_p,c-1) + \rangeCount(1,i-1,c,c)$$
where $c = \tBWT[i]$.
We concentrate on the scenario when $\tBWT[i] \in [-\sigma_p, \sigma_p]$.

\begin{figure}[!t]
\begin{center}
	\begin{subfigure}[t]{0.45\textwidth}
    	\centering
		\includegraphics[scale=0.8]{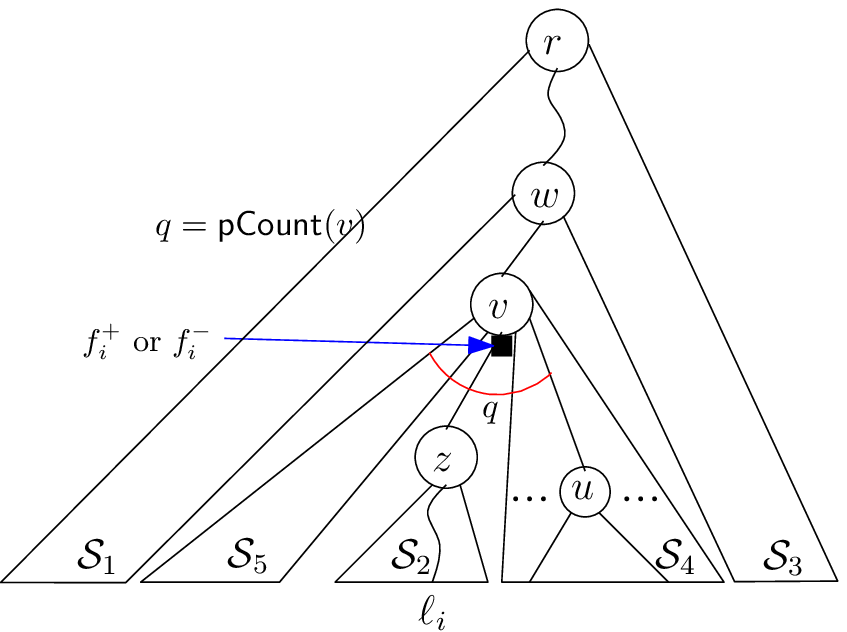}
		\caption{$f_i^+$ or $f_i^- = |\nodePath(v)|+1$} 
		\label{fig:leadingLFStruct}
	\end{subfigure}
	\hspace{2mm}
	\begin{subfigure}[t]{0.45\textwidth}
    	\centering
		\includegraphics[scale=0.8]{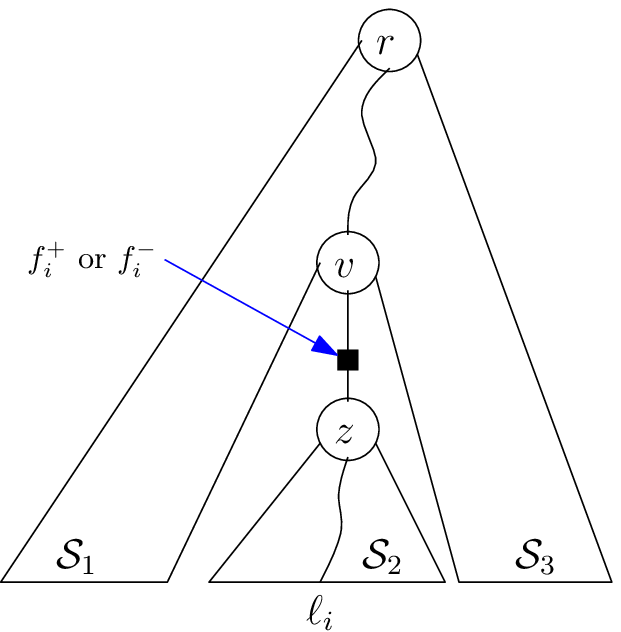}
		\caption{$f_i^+$ or $f_i^- > |\nodePath(v)|+1$}
		\label{fig:nonLeadingLFStruct} 
	\end{subfigure}
\end{center}
\caption{Illustration of various suffix ranges when the suffix $\T_{\tSA[i]}$ is preceded by a p-character}
\label{fig:leadNonLeadLFStruct}
\end{figure}

The following remain unchanged: (a) $\leafc(j)$ for a leaf $\ell_j$, and (b) $\pc(x)$ for a node $x$.
First find $z = \zeroNodePos(\ell_i)$ or $z = \zeroNodeNeg(\ell_i)$ depending on whether $\pBWT[i] \in [1,\sigma_p]$ or $\pBWT[i] < 0$.
Now, locate the node $v = \parent(z)$.
Depending on whether $\leafc(i) = 0$, or not, we find the desired ranges $\s_1$, $\s_2$, $\s_3$, and if required $\s_4$ and $\s_5$.
(See Figure~\ref{fig:leadNonLeadLFStruct} for illustration.)
Define $N_k$ to be the number of leaves $\ell_j$ in $\s_k$ such that $\tLF(j) \leq \tLF(i)$.
Then, $\tLF(i) = N_1+N_2+N_3+N_4+N_5$ is computed as follows.

\subsubsection{Computing $N_1$} 

%
%
%

For any $\ell_j \in \s_1$, $\tLF(j) < \tLF(i)$ iff one of the following holds:
\begin{itemize}
\item $\tBWT[j] \in [1,\sigma_p]$ and $f_j^+ > 1+|\nodePath(\lca(z,\ell_j))|$
\item $\tBWT[j] < 0$ 
\end{itemize}
%
%
Therefore, 
\begin{equation*}
N_1 =
\begin{cases}
\fsPos(v) + \rangeCount(1, L_v-1, -\sigma_p, -1), & \text{if } \leafc(i) = 0,
\\
\fsPos(z) + \rangeCount(1, L_z-1, -\sigma_p, -1), &  \text{otherwise}. 
\end{cases}
\end{equation*}

\subsubsection{Computing $N_2$}

\begin{enumerate}

\item
Assume $\tBWT[i] \in [1,\sigma_p]$.
For any leaf $\ell_j \in \s_2$, $\tLF(j) \leq \tLF(i)$ iff  one of the following holds:
\begin{itemize}
\item $\tBWT[j] \in [1,\sigma_p]$ and either $\tBWT[j] > \tBWT[i]$ or $\tBWT[j] = \tBWT[i]$ and $j \leq i$
\item $\tBWT[j]<0$
\end{itemize}
Therefore, 
$$N_2 = \rangeCount(L_z, R_z,c+1,\sigma_p) + \rangeCount(L_z, i,c,c) + \rangeCount(L_z, R_z,-\sigma_p,-1)$$ 
where $c = \tBWT[i]$.

\item
Assume $\tBWT[i] < 0$.
For any leaf $\ell_j \in \s_2$, $\tLF(j) \leq \tLF(i)$ iff one of the following holds:
\begin{itemize}
\item $-1 \geq \tBWT[j] > \tBWT[i]$
\item $\tBWT[j] = \tBWT[i]$ and $j \leq i$
\end{itemize}
Therefore, $N_2 = \rangeCount(L_z, R_z,c+1,-1)+\rangeCount(L_z, i,c,c)$, where $c = \tBWT[i]$.

\end{enumerate}

\subsubsection{Computing $N_3$}

For any leaf $\ell_j \in \s_3$, $\tLF(j) < \tLF(i)$ iff $\pBWT[j] < 0$ and $f_j^- \leq 1+|\nodePath(\lca(z,\ell_j))|$.
Therefore, 
%
\begin{equation*}
N_3 =
\begin{cases}
\rangeCount(1, R_v + 1, -\sigma_p, -1) - \fsNegRev(v), & \text{if } \leafc(i) = 0,
\\
\rangeCount(1, R_z + 1, -\sigma_p, -1) - \fsNegRev(z),&  \text{otherwise}. 
\end{cases}
\end{equation*}

%
%

\subsubsection{Computing $N_4$}
Let $u$ be the $\pc(v)$th child of $v$. 
Note that for any leaf $\ell_j \in \s_4$ such that $\tBWT[j] \in [1, \sigma_p]$, $f_j^+ \neq |\nodePath(v)|+1$; otherwise, the suffix $j$ should not have deviated from $i$ at the node $v$.
Likewise, if $\tBWT[j] < 0$, then $f_j^- \neq |\nodePath(v)|+1$.

\begin{enumerate}

\item 
Assume $\tBWT[i] > 0$.
Then, $N_4$ is the number of leaves $\ell_j$ in $\s_4$ satisfying $j \leq R_u$ and one of the following:
\begin{itemize}
\item $\sigma_p \geq \tBWT[j] \geq \tBWT[i]$
\item $\tBWT[j] < 0$
\end{itemize}
Therefore, $N_4 = \rangeCount(R_z+1, R_u,\tBWT[i],\sigma_p) + \rangeCount(R_z+1, R_u,-\sigma_p, - 1)$.

\item 
Assume $\tBWT[i] < 0$.
Then, $N_4$ is the number of leaves $\ell_j$ in $\s_4$ such that $j \leq R_u$ and $-1 \geq \tBWT[j] > \tBWT[i]$ i.e., $N_4 = \rangeCount(R_z+1, R_u,\tBWT[i] - 1,-1)$.
\end{enumerate}

\subsubsection{Computing $N_5$}
Note that for any leaf $\ell_j \in \s_5$ such that $\tBWT[j] \in [1, \sigma_p]$, $f_j^+ \neq |\nodePath(v)|+1$; otherwise, the suffix $j$ should not have deviated from $i$ at the node $v$.
Likewise, if $\tBWT[j] < 0$, then $f_j^- \neq |\nodePath(v)|+1$.
Also, note that the leading character of the path from $v$ to $\ell_j$ is negative.

\begin{enumerate}

\item 
Assume $\tBWT[i] > 0$.
Then, $N_5$ is the number of leaves $\ell_j$ in $\s_5$ that satisfies one of the following:
\begin{itemize}
\item $\sigma_p \geq \tBWT[j] \geq \tBWT[i]$
\item $\tBWT[j] < 0$
\end{itemize}
Therefore, $N_5 = \rangeCount(L_v,L_z-1,\tBWT[i],\sigma_p) + \rangeCount(L_v,L_z-1,-\sigma_p, - 1)$.

\item 
Assume $\tBWT[i] < 0$.
Then, $N_5$ is the number of leaves $\ell_j$ in $\s_5$ such that $-1 \geq \tBWT[j] > \tBWT[i]$ i.e., $N_5 = \rangeCount(L_v,L_z-1,\tBWT[i] + 1,-1)$.
\end{enumerate}
\paragraph{•}
Summarizing, we obtain the following result.

\BT
\label{thm:tLF}
We can compute $\tLF(i)$ as follows: \emph{(a)} in $\O(\log \sigma)$ time using $\O(n\log\sigma)$ bits, and \emph{(b)} in $\O(\log \sigma \log \log \sigma)$ time using $n\log\sigma+\O(n)$ bits.
\ET

\subsection{Backward Search}
As in Section~\ref{sec:backwardSearch}, assume that we have already found the suffix range $[\sp_1,\ep_1]$ of $\compl(Q)$.
We have to find the suffix range of $\compl(cQ)$.
If $c$ is a static character, then 
\begin{align*}
\sp_2 &= 1 + \rangeCount(1,n,-\sigma_p,c-1) + \rangeCount(1,\sp_1-1,c,c)\\
\ep_2 &= \rangeCount(1,n,-\sigma_p,c-1) + \rangeCount(1,\ep_1,c,c)
\end{align*}
Now, we consider the scenario when $c$ is a parameterized character.
We omit most of the details as they are similar to those in Section~\ref{sec:backwardSearch}.
We have the following two cases.

\subsubsection{Case 1 (Neither $c$ nor its complement appears in $Q$)}
Let $d$ be the number of distinct p-characters in $Q$, which can be computed in $\O(1)$ time.
Note that $\tLF(i) \in [\sp_2,\ep_2]$ iff $i \in [\sp_1,\ep_1]$, and either (a) $\tBWT[i] \in [1,\sigma_p]$ and $f_i^+ > |Q|$, or (b) $\tBWT[i] < 0$ and $f_i^- > |Q|$.
Then, 
$$(\ep_2 - \sp_2 + 1) = \rangeCount(\sp_1,\ep_1, d+1,\sigma_p) + \rangeCount(\sp_1,\ep_1,-\sigma_p,-d-1)$$
Let $u = \lca(\ell_{\sp_1},\ell_{\ep_1})$.
For any $i$, $\tLF(i) < \sp_1$ iff one of the following conditions hold:
\begin{itemize}
\item $i < \sp_1$, $\tBWT[i] \in [1,\sigma_p]$ and $f^+_i > 1 + |\nodePath(\lca(u,\ell_i))|$
\item $i \in [\sp_1,\ep_1]$, $\tBWT[i] < 0$ and $f^-_i \leq |Q|$
\item $i < \sp_1$ and $\tBWT[i] < 0$
\item $i > \ep_1$, $\tBWT[i] < 0$ and $f^-_i \leq 1 + |\nodePath(\lca(u,\ell_i))|$
\end{itemize}
Therefore, 
\begin{align*}
\tLF(i) &= 1 + \fsPos(u) + \rangeCount(\sp_1,\ep_1,-d,-1) + \rangeCount(1,\sp_1-1,-\sigma_p,-1)\\ 
&+ \rangeCount(\ep_1+1,n,-\sigma_p,-1) - \fsNegRev(u)
\end{align*}

\subsubsection{Case 2 ($c$ or its complement appears in $Q$)}
Assume that the number of characters until the first occurrence of $c$ (resp. $c$'s complement) in $Q$ is $f^+$ (resp. $f^-$).
If $f^+$ or $f^-$ does not exist, we take it to be $|Q| + 1$.
Let $d^+$ and $d^-$ be respectively the number of distinct p-characters in $Q[1,f^+]$ and $Q[1,f^-]$ respectively.
It follows from the arguments in Section~\ref{sec:backwardSearchCase2} that $d^+$ and $d^-$ can retrieved in $\O(1)$ time after an initial $\O(|P|\log \sigma)$ time pre-processing.

\begin{itemize}
\item \textbf{Case when $f^+ < f^-$:}
Note that $\tLF(i) \in [\sp_2,\ep_2]$ iff $i \in [\sp_1,\ep_1]$, $\tBWT[i] \in [1,\sigma_p]$ and $f_i = f^+$.
Consider any $i,j \in [\sp_1,\ep_1]$ such that $i<j$, both $\tLF(i), \tLF(j) \in [\sp_2,\ep_2]$, and both $\tBWT[i], \tBWT[j] \in [1,\sigma_p]$.
Now, $f^+_i = f^+_j = f^+$, and $\tLF(i) < \tLF(j)$.
Therefore, 
\begin{align*}
(\ep_2-\sp_2+1) &= \rangeCount(\sp_1,\ep_1,d^+,d^+) \text{, and } \\
\sp_2 &= \tLF(\min\{j \mid j \in [\sp_1,\ep_1] \text{ and } \tBWT[j] = d^+\})
\end{align*}

\item \textbf{Case when $f^+ > f^-$:}
Based on the above arguments, we can similarly derive the following.
\begin{align*}
(\ep_2-\sp_2+1) &= \rangeCount(\sp_1,\ep_1,-d^-,-d^-) \text{, and } \\
\sp_2 &= \tLF(\min\{j \mid j \in [\sp_1,\ep_1] \text{ and } \tBWT[j] = -d^-\})
\end{align*}
\end{itemize}
Finally, we remark that we can suitably modify Lemma~\ref{lem:saiSA} in a rather straight-forward way, so as to find the $\tSA[j]$ of any $j$ belonging to the suffix range of $\compl(P)$.
The time and space requirements remain unchanged.
By appropriately choosing $\Delta$, we arrive at Theorem~\ref{thm:structural}.

\section{Dictionary Matching}
\label{sec:dictionary}
Idury and Sch{\"{a}}ffer considered the following variant of Problem~\ref{problem}, known as \emph{Parameterized Dictionary Matching}.

\BDE[\cite{IduryS94}]
\label{prob:dict}
Let $\Sigma$ be an alphabet of size $\sigma \geq 2$, which is the union of two disjoint sets: $\Sigma_s$ having $\sigma_s$ static characters (s-characters) and $\Sigma_p$ having $\sigma_p$ parameterized characters (p-characters).  
Let $\calD$ be a collection of $d$ patterns $\{\P_1, \P_2, \dots, \P_d\}$ of total length $n$ characters that are chosen from $\Sigma$. 
The task is to index $\calD$, such that given a text $T$ (also over $\Sigma$), we can report all the positions $j$ in $T$ where at least one pattern $\P_i \in \calD$ is a p-match with $T[j-|\P_i|+1,j]$.
\EDE
\BC
\label{convention:dict}
Without loss of generality, assume that no two patterns $\P_i$ and $\P_j$ exist such that $\prev(\P_i) = \prev(\P_j)$.
Furthermore, assume that $i < j$ iff $\prevRev(\P_i)$ precedes $\prevRev(\P_j)$ in the lexicographic order, where $\prevRev(\P) = \prev(\P[p]\circ \P[p-1] \circ \dots \circ \P[1])$ and $p = |\P|$.
\EC
Idury and Sch{\"{a}}ffer presented an index, based on the Aho-Corasick (AC) automaton~\cite{AhoC75}. 
The index occupies $\Theta(m \log m)$ bits, where $m \leq n+1$ is the number of states in the automaton, and can report all $occ$ occurrences in time $\O(|T|\log \sigma + occ)$.
Typically, an occurrence implies a pair: the position in the text, as well as a pattern in the dictionary that is a p-match with the same-length substring ending at this position.
The following theorem summarizes our contribution.
\BT
\label{thm:dict}
All $occ$ pairs $\langle j, \P_i \rangle$, where a pattern $\P_i \in \calD$ is a p-match with $T[j-|\P_i|+1,j]$ can be found as follows: 
\begin{enumerate}[label={\emph{(\alph*)}},ref={\thethm\alph*}]
\item in $\O(|T|\log \sigma + occ)$ time using an $\O(m\log \sigma + d\log(n/d))$-bit index.
\item in $\O(|T| \log \sigma \log \log \sigma + occ)$ time using an $m\log \sigma + \O(m + d\log (n/d))$-bit index.
\end{enumerate}
\ET

\subsection{Idury and Sch{\"{a}}ffer's Index}
Let us first look at the index of Idury and Sch{\"{a}}ffer~\cite{IduryS94}. 
We begin by obtaining $\prev(\P_i)$ for every $\P_i$ in $\calD$, and then create a trie for all the encoded patterns.
The number of nodes in the trie is $m \leq n+1$.
For each node $u$ in the trie, denote by $\nodePath(u)$ the string formed by concatenating the edge labels from root to $u$.
Mark a node $u$ in the trie as \emph{final} iff $\nodePath(u) = \prev(\P_i)$ for some $\P_i$ in $\calD$.
Clearly, the number of final nodes is $d$.

For any $\prev$-encoded string $\prev(S)$ of a string $S$, and an integer $j \in [1,|S|]$, we obtain a string $\zeta(S,j)$ as follows.
Initialize $\zeta(S,j) = \prev(S)[j,|S|]$. 
For each $j' \in [1,|S|-j+1]$, assign $\zeta(S,j)[j'] = 0$ iff $\zeta(S,j)[j'] \geq j'$ and $\zeta(S,j)[j'] \notin \Sigma_s$.
Conceptually, $\zeta(S,j) = \prev(S[j,|S|])$.

Each node $u$ is associated with $3$ links as defined below:
\begin{itemize}
\item $\nxt(u,c) = v$ iff the label on the edge from the node $u$ to $v$ is labeled by the character $c$. 
This transition can be carried out in $\O(1)$ time using a perfect hash function.
\item $\failure(u) = v$ iff $\nodePath(v) = \zeta(\nodePath(u),j)$, where $j > 1$ is the smallest index for which such a node $v$ exists.
If no such $j$ exists, then $\failure(u)$ points to the root node.
Conceptually, this represents the smallest shift to be performed in $T$ in case of a mismatch.
\item $\report(u) = v$ iff $v$ is a final node and $\nodePath(v) = \zeta(\nodePath(u),j)$, where $j > 1$ is the smallest index for which such a node $v$ exists.
If no such $j$ exists, then $\report(u)$ points to the root node.
Conceptually, this represents a pattern which has an occurrence ending at the current symbol of the text.

\end{itemize}
Summarizing, the total space needed by the index is $\Theta(m \log m)$-bits.
Moving forward, we use the terms node and state interchangeably.
Likewise, for links and transitions.

To find the desired occurrences, first obtain $T' = \prev(T)$.
Now, match $T'$ in the trie as follows.
Suppose, we are considering the position $j$ in $T'$ (initially, $j=1$), and we are at a node $u$ with $|\nodePath(u)| = \delta_u$ i.e., we have matched $T[j,j+\delta_u-1]$ in the trie. 
First, repeatedly follow $\report$-links starting from $u$ until the root node is reached. 
Effectively, we report all patterns with a p-match ending at $(j+\delta_u-1)$.
Now, look at the character $c = T'[j+\delta_u]$ to match.
If $T'[j+\delta_u] \geq \delta_u$ and $T'[j+\delta_u] \notin \Sigma_s$, then take $c = 0$.
If $\nxt(u,c)$ is defined then follow it to a node $v$, and try to match an outgoing edge of $v$ with the character $T'[j+\delta_v]$.
Otherwise, follow $\failure(u)$ to a node $w$.
Now, we try to match $c$ with an outgoing edge of $w$; however, we are now considering the position $j = j+|\nodePath(u)| - |\nodePath(w)|$.
Repeat this process until the last character of $T'$ is reached.
The total time required is  $\O(|T|\log \sigma + occ)$\footnote{
Initially $T'$ is obtained in $\O(|T|\log \sigma)$ time.
On following a $\report$ link, either we report an occurrence, or we reach the root.
Following this, either we take a $\nxt$ transition or we follow a $\failure$ link; the number of such operations combined is at most  $2|T|$. 
Since each transition takes $\O(1)$ time, the total time is $\O(|T|\log \sigma + occ)$.
}.

\subsection{Representing States}
Belazzougui~\cite{Belazzougui10} obtained a succinct representation of the AC automaton for the classical dictionary indexing problem~\cite{AhoC75}.
Among other techniques, the central idea is to use a Succinctly Indexable Dictionary (SID);  see Fact~\ref{fact:SID} below.
For a state $u$, let $\nodePathRev(u)$ denote the reverse of $\nodePath(u)$.
A state $u$ is labeled by $r_u$ i.e., the lexicographic rank of $\nodePathRev(u)$ among the set $\{\nodePathRev(v) \mid v \text{ is a state in the AC automaton}\}$.
A transition $\nxt(u,c)$ is depicted by the bit representation of the string $c \circ r_u$.
By maintaining all representations in an SID, the desired node for a transition $\nxt(u,c)$ can be computed in $\O(1)$ time.
The space needed is $m \log \sigma + \O(m)$ bits.
\BF[\cite{RamanRR02,RamanRS07}]
\label{fact:SID}
A set $\cal S$ of $k$ integer keys from a universe of size $U$ can be stored in $k (\log (U/k) + \O(1))$ bits of space to support the following two operations in $\O(1)$ time.
\begin{itemize}
\item return the key of rank $i$ in the natural order of integers
\item return the rank of key $j$ in the natural order of integers if $j \in \cal S$. 
Otherwise, return $-1$.
\end{itemize}
\EF
An important observation related to the index of Idury and Sch{\"{a}}ffer~\cite{IduryS94}, described in the previous section, is that the edges in the trie are labeled from an alphabet of size $\Theta(m)$ in the worse case scenario.
This proves to be the primary bottleneck that prevents us from directly applying the technique of Belazzougui~\cite{Belazzougui10} to obtain a succinct representation. 
More specifically, the SID in this case will need $m\log m + \O(m)$ bits of space, which is not desirable.

To alleviate this problem, we first modify the labeling in the trie of Idury and Sch{\"{a}}ffer.
Moving forward, we denote the trie by $\calT$.
Assign a bit to every edge, which is set to $1$ iff the labeling on this edge corresponds to a p-character.
This can be easily achieved while building $\calT$.
Accordingly, we categorize an edge as a \emph{p-edge} or an \emph{s-edge}.
Every p-edge is in one of two states: \emph{visited} or \emph{new}.
Initally all p-edges are new.
Initialize a counter $C = 1$.
Traverse from the root node to the leftmost leaf (the ordering of the leaves can be in arbitrary order), and modify the labeling as follows:
\begin{itemize}
\item 
For an s-edge $(u,v)$, store the value of $C$ at $v$, and move to the next edge on the path.

\item 
If a p-edge $(u,v)$ is labeled by $0$, then label it by the current value of $C$. 
Store the value of $C$ at $v$, and then increment $C$ by $1$.
Move to the next edge on the path.

\item 
If a p-edge $(u,v)$ is labeled by $c>0$, then assign it the same label as the $(c+1)$th edge on the path from $v$ to root.
Store the value of $C$ at $v$, and move to the next edge on the path.
\end{itemize}
After the $i$th leftmost leaf $\ell_i$ is reached, find $x = \lca(\ell_i,\ell_{i+1})$, and use the stored value of $C$ at $x$ to label the edges from $x$ to $\ell_{i+1}$.
By pre-processing the trie with the data structure of Section~\ref{sec:treeTopology}, the entire process can be carried out in $\O(m)$ time.

Observe that each edge is now labeled by a character from the $\Sigma$.
Let $\nodePathRev_\Sigma(u)$ (resp. $\nodePath_\Sigma(u)$) denote the concatenation of the new edge labels on the path from $u$ to root (resp. from root to $u$).
With slight abuse of notation, let $\prevRev(u) = \prev(\nodePathRev_\Sigma(u))$.
Each state $u$ is \emph{conceptually labeled} by the lexicographic rank of $\prevRev(u)$ in the set $\{\prevRev(v) \mid v \text{ is a node in the trie}\}$.
Thus, each state is labeled by a number in $[1,m]$, where the root is labeled by $1$.
To distinguish the final states, we \emph{explicitly store} their labels in a SID; space required is $d\log (m/d) + \O(d)$ bits.
Using this, given the label of a final state, we can find the corresponding pattern in $\O(1)$ time (refer to Convention~\ref{convention:dict} and Fact~\ref{fact:SID}).
Lastly, we maintain a bit-vector $L[1,m]$ such that $L[j] = 1$ iff the state with label $j$ is a leaf in the trie.
Total space occupied is $m + d\log (m/d) + \O(d)$ bits.
\subsection{Handling $\nxt$ transitions}
We begin with a few notations.
For any p-edge $e = (w,x)$, we define $\Z(x) = $ the number of $0$'s in $\prevRev(w)[1,f_x]$, where $f_x$ is the first occurrence of the p-character labeling $e$ in the string $\nodePathRev_\Sigma(w)$.
If, $f_x$ is not defined, we let $\Z(x)$ equal the number of $0$'s in $\prevRev(x)$.
For any s-edge $e = (w,x)$, we define $\Z(x) = $ the label of the edge $e$.
As before, we map the s-characters to the interval $[\sigma_p + 1, \sigma]$, where the $i$th smallest s-character has value $(\sigma_p+i)$.

\BO
\label{obs:orderZ}
For any node $x$, $\Z(x_i) \neq \Z(x_j)$ for any two children $x_i$ and $x_j$ of $x$.
Also, $\prevRev(x_i)$ is lexicographically smaller than $\prevRev(x_j)$ iff at least one of the following holds: 
\begin{itemize}
\item 
$\sigma_p < \Z(x_i) < \Z(x_j)$
\item 
$\Z(x_i) \leq \sigma_p < \Z(x_j)$
\item 
$\sigma_p \leq  \Z(x_j) < \Z(x_i)$
\end{itemize}
\EO
%
\BO
\label{obs:matchT}
Suppose we are at a node $u$, and have matched $T[j, j+ \delta_u-1]$.
We have to select the edge corresponding to $c = T[j+\delta_u]$.
If $c \in \Sigma_s$, then clearly we need to select the s-edge $(u,v)$ such that $\Z(v) = c$.
Now, assume that $c \in \Sigma_p$.
\begin{itemize}
\item 
If $c \notin T[j, j+ \delta_u-1]$, then clearly we need to select the edge $e = (u,v)$ such that $\Z(v)$ equals the number of distinct p-characters in $T[j, j+ \delta_u]$.
\item
Otherwise, let the number of distinct p-characters in $T[j',j+ \delta_u-1]$ be $z$, where $j'$ is the last occurrence of $c$ in $T[j, j+\delta_u-1]$.
Then, we have to select the edge $(u,v)$ such that $\Z(v) = z$.
\end{itemize}
\EO
For any two distinct nodes $u$ and $v$ in $\calT$, we denote $u \prec v$ iff $\prevRev(u)$ is lexicographically smaller than $\prevRev(v)$. 
Since $\prevRev(u) \neq \prevRev(v)$, the relation $u \prec v$ is well-defined.

We create a compressed $\calTRev$ as follows.
Initially $\calTRev$ is empty.
For each non-leaf node $u$ in $\calT$, we add the string $\prevRev(u)\circ \$_{u,i}$ to $\calTRev$ for each child $u_i$ of $u$.
Clearly, each string corresponds to a leaf, say $\ell_{u,i}$, in $\calTRev$. 
We order the leaves according to the (lexicographic) rank of the string they represent.
The (lexicographic) rank of two strings $\prevRev(u)\circ \$_{u,i}$ and $\prevRev(u)\circ \$_{u,j}$ (i.e., of two leaves which share the same parent) are determined according $\Z(u_i)$ and $\Z(u_j)$ as outlined in Observation~\ref{obs:orderZ}.
If two leaves have distinct parents, then their order is defined by the relation $\prec$ on their parents.
In both the cases, the rank of the two strings corresponding to any two leaves is well defined.

Note that the number leaves in $\calTRev$ is same as the number of edges in $\calT$.
Therefore, it has $(m-1)$ leaves and at most $(m-2)$ internal nodes.
Maintain the array $\Z$, corresponding to the leaves in $\calTRev$, as a Wavelet Tree. 
Also, maintain a succinct representation of $\calTRev$. 
The total space required to store this information is $m\log \sigma + \O(m)$ bits.
The following lemma is a direct consequence of Observation~\ref{obs:prevOcc}, Lemma~\ref{lem:atleastOneStatic}, and Observation~\ref{obs:orderZ}.
\BL 
\label{lem:dictOrder}
Consider two nodes $u$ and $v$ (not necessarily distinct) and its respective children $u_i$ and $v_j$ in $\calT$.
Let the respective characters (from $\Sigma$) on the edges be $c_i$ and $c_j$.
\begin{enumerate}[label={\emph{(\alph*)}}]
\item 
If $c_i$ is parameterized and $c_j$ is static, then $\nxt(u,c_i) \prec \nxt(v,c_j)$.
\item If both $c_i$ and $c_j$ are static, then $\nxt(u,c_i) \prec \nxt(v,c_j)$ iff one of the following holds:
\begin{itemize}
\item
 $\Z(u_i) < \Z(v_j)$  
\item  
$\Z(u_i) = \Z(v_j)$ and $u \prec v$. 
Note that in this case $u \neq v$ and $\prec$ is defined.
\end{itemize}
\item 
Assume both $c_i$ and $c_j$ are parameterized and either $u \prec v$ or $u = v$.
Let $x = \lca(u,v)$, and $z$ be the number of $0$'s in $\prev(x)$.
\begin{enumerate}[label={\emph{(\arabic*)}}]
\item
If $\Z(u_i), \Z(v_j) \leq z$, then $\nxt(u,c_i) \prec \nxt(v,c_j)$ iff $\Z(u_i) \geq \Z(v_j)$.
\item
If $\Z(u_i) \leq z < \Z(v_j)$, then $\nxt(v,c_j) \prec \nxt(u,c_i)$.
\item
If $\Z(v_j) \leq z < \Z(u_i)$, then $\nxt(u,c_i) \prec \nxt(v,c_j)$.
\item
If $\Z(u_i), \Z(v_j) > z$, then $\nxt(v,c_j) \prec \nxt(u,c_i)$ iff all of the following are true:
\begin{itemize}
\item $\Z(u_i) = z+1$.
\item the leading character on the path from $x$ to $\ell_{u,i}$ is $0$.
\item the leading character on the path from $x$ to $\ell_{v,j}$ is not an s-character.
\end{itemize}
\end{enumerate}
\end{enumerate}
\EL 
Crucially, Lemma~\ref{lem:dictOrder} above is exactly the same as Lemma~\ref{lem:atleastOneStatic} with the following changes:
\begin{itemize}
\item $\pBWT$ replaced by $\Z$.
\item $\pLF$ mapping replaced by $\nxt$ transition.
\item the $\pST$ replaced by the compressed trie $\calTRev$.
\end{itemize}
Moreover, recall that in the data structure to compute $\pLF$ mapping, we only use the property that $\pST$ is a compressed trie (i.e., marking scheme, and the structure for computing $\fs$ and $\zeroNode$ only use this property.)
We simply redefine $\alphaDepth(\ell_{u,i})$ corresponding to a leaf $\ell_{u,i}$ as the highest node $v$ on the root to $\ell_{u,i}$ path that satisfies $\alphaDepth(v) \geq \Z(u_i)$.

Summarizing, in order to compute the $\nxt$ transitions, we construct a structure similar to that for computing $\pLF$ mapping. 
Using this, we can compute a $\nxt(u,c)$ transition in $\calT$ in the same space-and-time trade-off, provided that we know the corresponding leaf and $\Z$ entry in $\calTRev$.
To this end, we maintain a binary string $S$ as follows.
Traverse the leaves from left to right in $\calTRev$, and append a $1$-bit for each leaf, followed by a $0$-bit whenever we switch from a leaf to the next which has a different parent. 
Append a $0$ both at the beginning and at the end, and maintain a $\rank$-$\select$ structure.
The space needed is $\O(m)$ bits.

Suppose we are at a node $u$ in $\calT$, and have matched $T[j,j+\delta_u-1]$, where $\delta_u = |\nodePath(u)|$. 
We have to find the edge corresponding to the character $c = T[j+\delta_u]$ i.e., find $\nxt(u,c)$.
If $u$ is a leaf (which can verified by the bit-array $L$), we know there is no next transition, and we will follow $\failure(u)$.
Otherwise, let $k$ be the rank of $u$.
Based on Observation~\ref{obs:matchT}, if $c$ is an s-character, then we use it directly; otherwise, we find the encoding $z$ corresponding to $c$ in amortized $\O(\log \sigma)$ time per each character in $T$\footnote{
We maintain an array $A$ of length $\sigma_p$ such that $A[c]$ equals the position of last occurrence of $c \in \Sigma_p$ in the current window $T[j,j+\delta_u-1]$ of the text.
Initially, for each $c \in \Sigma_p$, we assign $A[c] = -1$.
Maintain a balanced binary search tree (BST) $\calT_{bin}$ containing only the p-characters $c$ indexed by $A[c] \neq -1$. 
Note that the size of $\calT_{bin}$ is $\O(\sigma_p)$, which implies every deletion, insertion, and search operation requires $\O(\log \sigma_p)$ time.

Suppose we are looking at a p-character at position $k$ in the text. 
If $A[T[k]] = -1$, we add $T[k]$ to $\calT_{bin}$, and find the number of characters in $\calT_{bin}$, which gives us the desired encoding.
Otherwise, find the number of entries in $\calT_{bin}$ with key at least $A[T[k]]$, which gives us the desired encoding.
Finally, update $A[T[k]] = k$ and proceed.

When we follow a failure link as we may have to remove several characters from $\calT_{bin}$.
Suppose, after following a failure link, we are considering the window of the text starting from $j' > j$. 
Clearly, we have to remove thee characters $c$ from $\calT_{bin}$ for which $A[c] < j'$.
Set $A[c] = -1$ for these characters.
The total number of such deletion operations is at most $|T|$, yielding an amortized time complexity of $\O(\log \sigma_p)$ per character.
}.
Locate the leaves $\ell_\sp$ and $\ell_\ep$ in $\calTRev$, where $\sp = \rank(\select_S(k,0),1) + 1$, and $\ep = \rank(\select_S(k+1,0),1)$.
The desired leaf and $Z$ entry are obtained as follows.
Obtain $q = \select_\Z(1 + \rank_\Z(\sp-1,z),z)$ in $\O(\log \sigma)$ time using the WT on $\Z$.
If $q > \ep$, then no such leaf exists, and we take the failure link.
Otherwise, the desired leaf is given by $\child(\parent(\ell_\sp), q-\sp+1)$.
The desired entry in $\Z$ is $\Z[q]$.
To see correctness, observe that $\ell_\sp$ and $\ell_\ep$ corresponds to the two children $u_i$ and $u_j$ of $u$ such that $\prevRev(u_i)$ and $\prevRev(u_j)$ are respectively the lexicographically smallest and largest strings in the set $\{\prevRev(w) \mid w \text{ is a child of } u\}$.
The correctness follows from Observation~\ref{obs:orderZ}, and the way the leaves under a single node in $\calTRev$ are arranged.

Summarizing, we obtain the following lemma.
\BL
\label{lem:dictNext}
The $\nxt$ transition in the index of Idury and Sch{\"{a}}ffer\emph{~\cite{IduryS94}} can simulated as follows: \emph{(a)} in $\O(\log \sigma)$ amortized time per character by using $\O(m \log \sigma)$ bits, and \emph{(b)} in $\O(\log \sigma \log \log \sigma)$ amortized time per character by using $m \log \sigma + \O(m)$ bits.
\EL
\subsection{Handling $\failure$ and $\report$ transitions}
Note that for any two nodes $u$ and $v$, if $\failure(u) = v$, then it $\prevRev(v)$ is the longest prefix of $\prevRev(u)$ that appears in $\calT$.
Similar remarks hold for $\report(u) = v$, where $v$ is a final node.
Therefore, these behave exactly in the same manner as in the case of traditional pattern matching, and we can simply use the idea of Belazzougui to perform these transitions (see Sections 3.3 and 3.4 in~\cite{Belazzougui10}). 
\BL
\label{lem:dictFailReport}
The $\failure$ and $\report$ links in the index of Idury and Sch{\"{a}}ffer\emph{~\cite{IduryS94}} can simulated in $\O(1)$ time by using $\O(m + d\log (n/d))$ bits.
\EL

\paragraph{•}
Putting Lemmas~\ref{lem:dictNext} and~\ref{lem:dictFailReport} together, we obtain Theorem~\ref{thm:dict}.
\section{Discussion}
\label{sec:conclusion}


We leave two main questions unanswered, both related to the central problem of parameterized pattern matching.
The first one, concerning the space consumption, is ``Can we convert $\O(n)$ bits term to $o(n)$ in our space requirements?''.  
The second one is related to construction of the index.
In the case of classical full text indexes, designing linear time construction algorithm in compact space (i.e., the working space is $\O(n \log \sigma)$ bits) has been an active area as seen from the seminal work of Hon, Sadakane, and Sung~\cite{HonSS03} and the more recent work of Belazzougui~\cite{BelazzouguiSTOC14}.
It would be interesting to check for the existence of a (randomized) algorithm for constructing a compressed index for the parameterized matching problem that uses $\O(n \log \sigma)$ working space and attains the same bounds of the best-known algorithms for constructing p-suffix trees~\cite{ColeH03a,Kosaraju95}.


\newpage

\bibliographystyle{alpha}
\bibliography{References}

\end{document}